\documentclass[number, sort&compress, times, 3p, number]{elsarticle}

\usepackage[utf8]{inputenc}

\usepackage{graphicx}
\usepackage{dcolumn}
\usepackage{bm}

\usepackage{amssymb}
\usepackage{amsmath}

\usepackage[colorlinks=true,linkcolor=black, citecolor=black, urlcolor=black]{hyperref}
\usepackage[capitalise]{cleveref}

\usepackage{xcolor} 

\usepackage{tikz}

\usepackage{easyReview}

\usepackage{natbib}
\newcommand{\R}{\mathbb{R}}
\newcommand{\N}{\mathbb{N}}

\newcommand{\Pe}{\text{Pe}}

\newcommand{\textmatrix}[1]{\left(\begin{smallmatrix} #1 \end{smallmatrix}\right)}
\newcommand{\bigmatrix}[1]{\begin{pmatrix} #1 \end{pmatrix}}

\begin{document}

\title
{Numerical simulation and analysis of mixing enhancement due to chaotic advection using an adaptive approach for approximating the dilution index}

\author[1]{Carla Feistner}
\ead{carla.feistner@fau.de}

\author[1]{Mónica Basilio Hazas}
\ead{monica.basilio@fau.de}

\author[2]{Barbara Wohlmuth}%
\ead{wohlmuth@cit.tum.de}

\author[1]{Gabriele Chiogna\corref{cor1}}%
\ead{gabriele.chiogna@fau.de}

\affiliation[1]{organization={Chair of Applied Geology, GeoZentrum Nordbayern, Friedrich-Alexander-Universität Erlangen-Nürnberg},
            addressline={Schlossgarten 5},
            postcode={91054},
            city={Erlangen},
            country={Germany}}
            
\affiliation[2]{organization={Chair of Numerical Mathematics, School of Computation, Information and Technology, Technical University of Munich},
addressline={Arcisstrasse 21},
postcode={80333},
city={Munich},
country={Germany}}

\cortext[cor1]{Corresponding author}

\date{November 13, 2024}

\begin{abstract}
  Lagrangian particle-tracking methods are particularly suitable to study solute transport in velocity fields displaying chaotic advection. They can accurately resolve stretching and folding processes, the increase in the solute-solvent interface available for diffusion as well as Kolmogorov--Arnold--Moser (KAM) islands, non-mixing regions that limit the chaotic area in the domain and, thereby, the mixing enhancement. However, they also display limitations due to the finite number of discrete particles, particularly if we are interested in the quantification of mixing processes, which require an accurate description of the particle density or concentration gradients. In this work, we use the dilution index to quantify the temporal increase in mixing of a solute within its solvent. We introduce a new approach to select a suitable grid size for the approximation of the density function, motivated by the theory of representative elementary volumes. It preserves the central feature of the dilution index, which is monotonically increasing in time, highlighting the importance of a suitable choice for the grid size in the dilution index approximation. We use this approach to demonstrate the mixing enhancement for two chaotic injection-extraction systems that exhibit chaotic structures: a source-sink dipole and a rotated potential mixing. Using our new approach, we assess the choice of design parameters of the injection-extraction systems to effectively engineer chaotic mixing. We demonstrate the important role of diffusion in filling the KAM islands and reaching complete mixing and, consequently, the importance of avoiding numerical diffusion, which often affects Eulerian methods applied on the advection-diffusion equation.
\end{abstract}


\maketitle

\section{Introduction}

The natural mixing processes in groundwater and microfluidic systems occurring by molecular diffusion are too slow for many applications \cite{stone_2001,stremler_2004,rolle_2019}.
However, advection can boost mixing processes between solute and solvent. This is particularly evident when the flow leads to deformations of the plume, as with shear flow, focusing and defocusing flow, twisting flow \cite{rolle_2019,paster_2015}, or helical flow \cite{ye_2015,ye_2016}.
It is important to note that while turbulent flows can enhance mixing \cite{warhaft_2000, kadoch_2012}, they are infeasible in certain situations, such as laminar flow in porous media. This limitation underscores the need for alternative methods. 
One of such alternatives is chaotic advection, a method that has shown potential to enhance fluid mixing \cite{stremler_2004,liu_2000,aref_1984} and can also occur in flows at low Reynolds numbers \cite{aref_1990}. 
Due to the stretching and folding, the surface between solute and solvent increases, and concentration gradients steepen, boosting the diffusion appearing over this surface \cite{ottino_1989,borgne_2014,weeks_1998}. The process of engineering chaotic flow in a system, which requires increased mixing, is called "design for chaos" \cite{stremler_2004,liu_2000}.

Mixers within laminar flow regimes have a wide range of applications. Beginning with the topic of groundwater remediation,
it is shown theoretically \cite{lester_2013}, numerically \cite{turuban_2018} and experimentally \cite{souzy_2020, heyman_2020,kree_2017} that the flow through porous media already exhibits chaotic trajectories at the pore scale. At the Darcy scale, the work of \cite{bagtzoglou_2007} gives theoretical evidence that engineered chaotic advection using randomly oscillating wells enhances groundwater remediation if successfully implemented into groundwater flow systems. In practice, an implementation was achieved in \cite{cho_2019} where the rotated potential mixing (RPM) flow was engineered at field scale, leading to improved lateral tracer spreading. 
Moreover, for applications concerning reactive transport the work of \cite{paster_2014} highlights that mixing serves as a limiting factor for reactions in diffusion-limited systems, indicating that chaotic mixing can be utilized in a variety of contexts involving reactions.
Another application is in microfluidics, where many processes are constrained by incomplete mixing in a laminar flow regime of low Reynolds numbers and high Péclet numbers \cite{aref_2017, ward_2015}. There are two main approaches to engineering mixing in these applications: passive and active. A range of passive mixing designs, such as flow-through mixers \cite{stremler_2004, liu_2000,song_2003} which are based on a given channel geometry, can be used to increase mixing involving sensitive species \cite{aref_2017}. These mixers find application in the synthesis of biomolecules, drug delivery, and diagnostic testing \cite{teh_2008}. 
In active mixers, including microstirrers, acoustic mixers, and flow pulsation \cite{ward_2015}, chaotic advection is designed using external perturbation sources \cite{aref_2017}.
A successful implementation of active mixing using pulsating flow was achieved in \cite{stremler_2004} for enhancing gene expression profiling.
Further applications of mixers range from food processing \cite{metcalfe_2009} and heat transfer devices \cite{aref_2017,ganesan_1997} to the understanding of chaotic flow through blood vessels \cite{schelin_2009}, showing that chaotic advection and the thereby increased degree of dilution is of high interest.

Modeling transport processes requires solving the classical advection-diffusion equation \cite{kitanidis_1994,basiliohazas_2022}. Besides the more classical Eulerian methods, like finite differences, finite elements, or finite volumes, one can approximate the solution to this equation using Lagrangian particle tracking. This approach is widely applied in the field of porous media for both conservative \cite{tartakovsky_2008, leborgne_2008,leborgne_2008a, srinivasan_2010} and reactive tracers \cite{tartakovsky_2010, paster_2015, rahbaralam_2015, paster_2014, morvillo_2021, salamon_2006}. In contrast to the Eulerian framework, the Lagrangian approach does not suffer from numerical dispersion or artificial oscillations, which requires the Eulerian approach to work with small grid cells and time steps for advection-dominated flows \cite{salamon_2006, schmidt_2017, younes_2005}. For conservative transport, the particle tracking approach also has the additional advantage of built-in mass conservation \cite{salamon_2006}.
For chaotic advection setups, the Lagrangian perspective reveals more information about the fractal plume structure \cite{bagtzoglou_2007} and is therefore used for this type of flow, among others, in \cite{aref_2002,bagtzoglou_2007}.
The passive advection of a particle is represented by the advection equation, an ordinary differential equation \cite{aref_1990}, while the diffusive process is added in form of a random walk \cite{bagtzoglou_2007, ghoniem_1985}.
For two-dimensional incompressible flows, the advection equation is defined by a Hamiltonian system \cite{bear_1988,aref_1990,aref_1984,chate_1999}.
The analysis of the resulting advective flow reveals basic properties.
Of interest may be Lyapunov exponents, as they show the strength of the stretching and thereby quantify the extent of chaotic flow structures. For volume-preserving flow, we will also find fixed points that are either elliptic or hyperbolic. They alternate on a chain of periodic points and thereby form structures known as island chains \cite{chate_1999}. 
If the flow is chaotic, we may observe the occurrence of Kolmogorov--Arnold--Moser (KAM) islands, which are nonmixing regions arising around the elliptic points in the flow \cite{lester_2009, chate_1999}. In the remaining chaotic region, the flow is guided by homoclinic or heteroclinic tangles of the manifolds belonging to unstable stagnation points, leading to exponential fast deformations \cite{chate_1999}.
Since KAM islands consist of quasi-periodic points \cite{luo_2006}, particles can only enter them by diffusion. Hence, they limit the chaotic area in the domain and decrease mixing \cite{chate_1999}.

The goal of stretching and folding is to enlarge the interface between solute and solvent, thereby increasing diffusive fluxes over this interface \cite{bagtzoglou_2007,neupauer_2014} and enhancing mixing processes. An approach to quantify the effect of mixing is the computation of the dilution index, an entropy-related approach to measure dilution \cite{kitanidis_1994}. The term dilution refers to the increase in volume of the fluid occupied by a solute. If the solute concentration follows the advection diffusion equation, this volume increases monotonically in time, hence one of the key properties of the dilution index is that it also increases monotonically in time \cite{kitanidis_1994}. 
For incompressible flow in a non-diffusive system, the volume of a conservative solute and hence also its dilution is constant in time, showing that the diffusion limits the increase in dilution.
This measure, applied on results of particle tracking simulations, has a strong dependence on the grid size. It is similar to the problem of estimating the concentration distribution from discrete particle locations that was also identified by \cite{rahbaralam_2015}. For a non suitable grid the approximation of the dilution index may not show monotonic growth in time, as we see in \cite{kapoor_1998}.

We aim to quantify the dilution increase due to two-dimensional chaotic flow structures and identify why different configurations can lead to significant differences in the mixing behavior. We investigate the evolution of a conservative, nonreactive, and nonsorbing plume inside the pulsed source-sink (PSS) system and the rotated potential mixing (RPM) flow. We model solute transport with Lagrangian random walk particle tracking, taking into account the time stepping, the parameters controlling the flow parametrization, and the diffusion coefficient to balance the Péclet number of the two systems. We characterize the flow by creating Poincaré sections revealing the KAM island structure and quantify mixing with the dilution index. 

Overall, the novelty of our work consists of the introduction of a numerical method to better estimate the dilution index for particle tracking simulations using adaptive grid size selection and the use of this method to evaluate the potentials and limitations of chaotic advection to enhance mixing.

Our paper is structured as follows. \cref{sec:math_theory} introduces the mathematical theory of the two chaotic advection setups and the random walk particle tracking. We also revisit the definition of the dilution index in \cref{sec:dil} and show some properties. In \cref{sec:methodology}, we concentrate on the numerical implementation of the chaotic system and consider the time stepping and Péclet numbers. In \cref{sec:dil_gaussian}, we use the example of a pulse injection to demonstrate the influence of the grid size, used to approximate the particle density distribution, on the resulting dilution index. The results motivate the deviation of the method to adaptively select a suitable grid size. Finally, in \cref{sec:num_experiments}, we show the results of our numerical experiments using our approach before giving a conclusion in \cref{sec:conclusion}.

\section{Mathematical theory}\label{sec:math_theory}

\subsection{Chaotic advection setups}\label{sec:CA_setups}

In this paper, we consider flow in two dimensions. 
Hence, to successfully engineer chaos, the advection equation must be unsteady \cite{ottino_1989,stremler_2004,aref_1990}.
By following a large number of tracer particles inside the system, we approximate the spatial concentration distribution of the plume over time.

\subsubsection{Pulsed source-sink system}\label{sec:PSS_math}

\begin{figure}[tbp!]
  \centering
  \includegraphics[width=8.6cm]{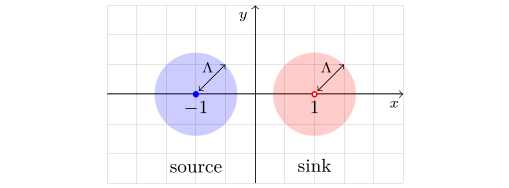}
  \caption{Two-dimensional plane with one sink and one source at $\pm\textmatrix{1\\0}$. All particles with a distance to the sink smaller than $\Lambda$ are extracted during the next stroke of the sink. Within the following stroke of the source, all of them are reinjected and occupy the circle around the source of radius $\Lambda$.}
\label{tikz:exp_setup}
\end{figure}

The first setup under consideration is called the pulsed source-sink system introduced in \cite{jones_1988} and later also addressed in \cite{sposito_1998, stremler_2004, sposito_2006}. We consider a two-dimensional plane with one sink and one source at arbitrary locations (see \cref{tikz:exp_setup}). During the experiment, the sink and the source are operated alternatively for a time interval of length $\tau$, meaning only one is switched on at a time. 
The particles extracted by the sink are reinjected by the source in the next stroke using the \textit{first-out-first-in} procedure \cite{stremler_2004}. This can be realized physically by collecting the extracted particle in a tube beneath the flow plane during the stoke of the sink. Prior to the stroke of the source the tube is flipped over and connected to the source where the fluid is then reinjected into the plane \cite{jones_1988}.

In general, the flow of an incompressible fluid in a non-deformable medium is described by a Lagrange stream function \cite{bear_1988,aref_1990,aref_1984,chate_1999}. The existence of a Lagrange stream function $\psi:\R^2 \rightarrow \R$ implies that the associated flow can be described using the Hamiltonian system \cite{sposito_1998,durst_2022}
\begin{align}\label{eq:PSS_hamiltonian_system}
  \frac{dx}{dt} = \frac{\partial \psi}{\partial y}, \qquad
  \frac{dy}{dt} = -\frac{\partial \psi}{\partial x}, \qquad
  \bigmatrix{x(0)\\y(0)} = \bigmatrix{x_0\\y_0}.
\end{align}
For a point source or sink at location $\textmatrix{\bar{x}\\\bar{y}}$, $\psi$ is given in \cite{sposito_2006} by
\begin{align*}
  \psi(x, y) = \frac{Q}{2\pi} \tan^{-1}\left(\frac{y-\bar{y}}{x-\bar{x}}\right)
\end{align*}
with $Q$ being the total areal discharge across any closed curve surrounding the source ($Q>0$) or the sink ($Q<0$).

The analytical solution of the flow defined in \eqref{eq:PSS_hamiltonian_system} is derived in \cite{sposito_2006,jones_1988}. We give a detailed derivation in \cref{app:PSS_analytical_sol}. Arguing that each positioning of sink and source can be transformed into an equivalent configuration with sink and source at $\pm\textmatrix{a\\0}$ and by multiplying the length scale by $1/a$, we assume sink and source at locations $\pm\textmatrix{1\\0}$. The flow is parametrized by a combination of $\tau$, $Q$, and $a$ in the definition of
\begin{align}\label{eq:Lambda}
  \Lambda(t) = \frac{1}{a} \sqrt{\frac{Q}{\pi}t}
\end{align}
with $\Lambda := \Lambda(\tau)$.
Starting a particle at location $\textmatrix{x\\y}$, the flow produced by the source ($\Psi_+$) or the sink ($\Psi_-$) is given by
\begin{align}\label{eq:PSS_advective_particle_motion}
  \Psi^t_\pm\bigmatrix{x\\y}  = 
  \bigmatrix{x \pm 1\\y} \,
  \sqrt{1 \pm \frac{\Lambda^2(t)}{(x \pm 1)^2 + y^2}} \mp \bigmatrix{1\\0}.
\end{align}

It holds that all particles within the circle of radius $\Lambda$ around the sink are extracted during the operation of the sink for time $\tau$. This circle is depicted as $B_\Lambda\textmatrix{1\\0}$ and also shown in \cref{tikz:exp_setup}.
For the extraction-reinjection procedure, the authors of \cite{jones_1988} derive the map $M_{\text{inj}}:B_\Lambda\textmatrix{1\\0} \rightarrow \R^2$, taking into consideration the \textit{first-out-first-in} reinjection protocol and an injection angle of $\pi - \theta$ given the extraction angle $\theta$. The map is given by
\begin{align*}
  M_{\text{inj}}\bigmatrix{x\\y} = \bigmatrix{1 - x\\y}\sqrt{\frac{\Lambda^2}{(x-1)^2+y^2} - 1} - \bigmatrix{1\\0}.
\end{align*}
Combining these expressions, we construct the Map $M:\R^2 \rightarrow \R^2$ representing the consecutive operation of source and sink where the sink is operated first:
\begin{align}\label{eq:PSS_M}
  M\bigmatrix{x\\y} = \begin{cases}
    \left(\Psi_+^{\tau} \circ \Psi_-^{\tau}\right)\textmatrix{x\\y} & \text{for } \textmatrix{x\\y} \in \R^2/B_\Lambda\textmatrix{1\\0} \\
    M_{\text{inj}}\textmatrix{x\\y} & \text{otherwise}.
  \end{cases}
\end{align}
Starting with the operation of the sink ensures that, after the application of $M$, a particle still has a defined location in $\R^2$.
Due to the definition of $\Lambda$, the parameters $a$, $Q$, and $\tau$ are redundant. 
The pulsed source-sink system is hence parametrized only by $\Lambda$.

\subsubsection{RPM flow}

As a second setup, we consider the rotated potential mixing (RPM) flow \cite{lester_2009}. The particle movement is produced using one source and one sink in a two-dimensional plane, operating nonstop and simultaneously. They are placed opposite each other on the edge of the unit circle around zero, which is our domain. The initial locations are $\textmatrix{0\\1}$ for the source and $\textmatrix{0\\-1}$ for the sink. After a time interval of length $\tau$ the source and the sink are rotated around the origin by the angle $\Theta$. Particles that are extracted by the sink are instantaneously injected by the source. 
Fixing source and sink at their initial positions, we can describe the flow using the Lagrange stream function $\psi$ 
\begin{align}\label{eq:RPM_psi}
  \psi(x, y) = \tan^{-1} \left(\frac{2x}{1-x^2-y^2}\right).
\end{align}
Like for the pulsed source-sink system, we can use the Hamiltonian system \eqref{eq:PSS_hamiltonian_system} with stream function \eqref{eq:RPM_psi} to describe the advective particle transport.

The analytical solution of the Hamiltonian system is derived in \cite{lester_2009}. We solve the flow in the reference system with source and sink at $\pm\textmatrix{0\\1}$. We give a detailed explanation in \cref{app:RPM_analytical}. Focusing first on the positive side of the $x$-axis (i.e., $x > 0$), we represent the particle position in terms of the angle to the origin $\theta \in (-\pi/2, \pi/2)$ and its streamline $\psi \in (0, \pi/2]$. This is convenient, as $\psi$ is constant in time. Using this coordinate system, the advection time of a particle along a streamline until it reaches $\theta = 0$, was established in \cite{lester_2009} as
\begin{align}\label{eq:RPM_t_adv}
  \begin{split}
    t_\text{adv}(\theta, \psi) = \csc^2(\psi) 
    \times\Biggl\{ &\cot(\psi) \arctan \left[ \frac{\sin(\theta) \cot(\psi)}{\sqrt{1 + \cos^2(\theta) \cot^2(\psi)}} \right] \\
    &+ \sin(\theta) \sqrt{1 + \cos^2(\theta) \cot^2(\psi)} 
    - |\cot(\psi)| (\theta + \cos(\theta) \sin(\theta)) \Biggr\}.
    \end{split}
\end{align}
The advection time yields a negative value for particles downstream of the centerline $\theta = 0$. The residence time of a particle on the streamline $\psi$ is hence given by
\begin{align}\label{eq:RPM_t_res}
  T_{\text{res}}(\psi) = \left(t_{\text{adv}}\left(\frac{\pi}{2}, \psi\right) - t_{\text{adv}}\left(-\frac{\pi}{2}, \psi\right)\right).
\end{align}
For computing the angle $\theta(t)$ of a particle $(\theta_0, \psi_0)$ after time $t$, we need to use the modulo operator $\bmod$ to account for the instantaneous reinjection of a particle. The formula describing the map $\theta_0 \mapsto \theta(t)$ is given by
\begin{align}\label{eq:RPM_formular_advection}
  \begin{split}
    t_{\text{adv}}(\theta(t), \psi_0) + \frac{T_{\text{res}}(\psi_0)}{2} =
    \left(t_{\text{adv}}(\theta_0, \psi_0) + \frac{T_{\text{res}}(\psi_0)}{2} - t\right) \mod T_{\text{res}}(\psi_0).
  \end{split}
\end{align}
Notice the correction of the sign in front of the $t$ on the right-hand side concerning equation (24) of \cite{lester_2009}. A more detailed explanation of the derivation of \eqref{eq:RPM_t_res} and \eqref{eq:RPM_formular_advection} is given in \cref{app:RPM_analytical}. 
Solving this equation yields the flow $\Psi_1^{t}$ for all particles on the positive side of the $x$-axis. For particles on the negative side, we use the reflection symmetry of $\psi$ with respect to the $y$-axis, yielding the flow $\Psi_2^{t}\textmatrix{x\\y} = \textmatrix{-1&0\\0&1}\Psi_1^{t}\textmatrix{-x\\y}$. 

The derivation in \cite{lester_2009} is missing the description for particles on the $y$-axis (i.e., $x = 0$) where the usage of $(\theta, \psi)$ coordinates is insufficient. Alternatively, we directly solve \eqref{eq:PSS_hamiltonian_system}
\begin{align}\label{eq:RPM_1D_ODE}
  \dfrac{d x}{d t} = \dfrac{4xy}{(y^2+x^2-1)^2+4x^2}, \qquad
  \dfrac{d y}{d t} = \dfrac{-2(x^2-y^2+1)}{(x^2+y^2)^2+2x^2-2y^2+1}, \qquad
  \bigmatrix{x(0)\\y(0)} = \bigmatrix{0\\y_0}
\end{align}
Here, we have used the definition of $\psi$ in \eqref{eq:RPM_psi}.
Given that $x(0) = 0$ we find $\frac{dx}{dt}|_{t=0} = 0$ and hence $x(t) = 0$ for all $t \in \R$. We can, therefore, simplify \eqref{eq:RPM_1D_ODE} to 
\begin{align*}
  x(t) = 0 , \qquad
  \dfrac{d y}{d t} = \dfrac{2}{y^2-1} , \qquad
  y(0) = y_0
\end{align*} 
and solve for $y$ by using the separation of variables
\begin{align*}
    \begin{split}
    y - \frac{1}{3} y^3 &= -2 t + y_0 - \frac{1}{3} y_0^3.
    \end{split}
\end{align*}
This leads to
\begin{align*}
  t_{\text{adv}}(y) = \frac{y}{2} - \frac{y^3}{6}, \qquad T_{\text{res}} = \frac{2}{3},
\end{align*}
which we can use to describe the particle motion on the $y$-axis with \eqref{eq:RPM_formular_advection} where $y$ takes the place of $\theta$, yielding the flow $\Psi_0^t$.
Combining the above flows, we can define the flow $\Psi^t:B_1\textmatrix{0\\0} \rightarrow B_1\textmatrix{0\\0}$ for the complete domain.

Extending the solution to the case where source and sink are rotated after time $\tau$, we introduce the rotation operator $R_k:\R^2\rightarrow\R^2$ and rotate the particles by the angle $k\Theta$. Using $R_{k}R_{l} = R_{k+l}$, the position of a particle $\textmatrix{x\\y}$ after time $n\tau$ is given by
\begin{align*}
  \bigmatrix{x(n\tau)\\y(n\tau)} = R_{n-1}\Psi^\tau R_{1-n} ... R_{1}\Psi^\tau R_{-1}R_{0}\Psi^\tau R_{0}\bigmatrix{x_0\\y_0}
  = R_{n}\left(R_{-1} \Psi^\tau\right)^n\bigmatrix{x_0\\y_0}.
\end{align*}
We define the map $M:B_1\textmatrix{0\\0}\rightarrow B_1\textmatrix{0\\0}$ representing the operation of the RPM flow for time $\tau$ by
\begin{align}\label{eq:RPM_M}
  M\bigmatrix{x\\y} = R_{-1} \Psi^\tau\bigmatrix{x\\y}.
\end{align}

\subsection{Random walk to model diffusion}\label{sec:random_walk}

Hamiltonian flows, as the ones defined above, only describe advective processes. 
To also simulate a diffusive process, we need to add some stochastic dynamics to the flow. Langevin and Einstein have shown how a random walk can be used to model the displacement of a Brownian particle \cite{lemons_1997,einstein_1906}. For our application, we assume a mean square displacement of $\sigma^2 t$ after time $t$ where $\sigma$ is the diffusion coefficient. 
Given the advective flow $\Psi^{\Delta t}$, the diffusion is added in terms of 
\begin{align}\label{eq:diffusive_basic_flow}
  \Psi^{\Delta t}_\sigma \bigmatrix{x\\y} = \Psi^{\Delta t}\bigmatrix{x\\y} + \sqrt{\Delta t}\, \sigma \bigmatrix{\xi_1\\\xi_2}, \quad \xi_1, \xi_2 \sim \mathcal{N}(0, 1).
\end{align}
It is shown in \cite{tome_2015} that the probability density of a particle following this modified flow does mirror the solution of the advection-diffusion equation with diffusion tensor $D = 1/2\textmatrix{\sigma^2&0\\0&\sigma^2}$.

\subsection{Dilution index}\label{sec:dil}

Our final goal is to analyze how well the injected tracer mixes within the system. Due to the stochastic nature of the flow $\Psi_\sigma$, defined in \eqref{eq:diffusive_basic_flow}, the position of a particle after time $t>0$ is not deterministic. It is described by a probability density function $\rho : \Omega \rightarrow \R$, which can also be seen as the normalized concentration distribution of a plume. 
Distinguishing between spreading and dilution is essential in this context, as only the latter is associated with an increase in the volume of fluid occupied by the solute. An approach for measuring the dilution of the tracer is the dilution index introduced by Kitanidis \cite{kitanidis_1994}. It is defined as the exponential over the differential entropy of $\rho$ over $\Omega$
\begin{align}\label{eq:E_continous}
  \mathcal{E} = \exp\left(- \int_\Omega \rho(x) \ln(\rho(x)) \, dx\right)
\end{align}
If the probability density function $\rho(x, t)$ represents a solution to the advection-diffusion equation with positive definite diffusion tensor, then $\frac{d \ln(\mathcal{E})}{dt} > 0$ \cite{kitanidis_1994}. 

For approximating \eqref{eq:E_continous}, we create a grid inside the domain $\Omega$ with $n$ grid cells. Each cell $k$ has a grid size $\Delta V_k$ and a mean particle probability density $\rho_k$. We can approximate the dilution index by replacing the integral with a finite sum:
\begin{align}\label{eq:E_approx}
  \begin{split}
    \mathcal{E} \approx \exp\left(- \sum_{k=1}^{n} \Delta V_k\, \rho_k \ln(\rho_k) \right).
  \end{split}
\end{align}
However, in our application, the exact representation of $\rho$ is not available. We compute an approximation by generating multiple realizations of the stochastic particle evolution defined in \eqref{eq:diffusive_basic_flow}. We estimate the probability $P_k$ of the particle to be contained in cell $k$ as the number of realizations contained in cell $k$ divided by the total amount of realizations. It follows $\rho_k \approx P_k/\Delta V_k$. 
The approximated dilution index on a nonuniform grid is therefore given by
\begin{align}\label{eq:E_nonuniform}
E = \exp\left(- \sum_{k=1}^{n} P_k \ln\left(\frac{P_k}{\Delta V_k}\right) \right).
\end{align} 
In case of an equidistant grid with grid size $\Delta V$, we can use $\sum_{k=1}^{\infty} P_k = 1$ to simplify \eqref{eq:E_nonuniform} to
\begin{align}\label{eq:E}
  \begin{split}
  E = \Delta V \exp\left(-\sum_{k=1}^n P_k \ln(P_k)\right).
\end{split}
\end{align} 
Looking at formula \eqref{eq:E}, we find the definition of the discrete entropy $H = -\sum_{k=1}^n P_k \ln(P_k)$ inside the exponential. It holds that
\begin{align*}
  0 \leq H \leq \ln(n)
\end{align*}
with $H = 0$ if and only if there exists one cell that contains all particles and $H = \ln(n)$ if and only if the particles are uniformly distributed over all cells \cite{bremaud_2017}. This directly implies
\begin{align}\label{eq:E_bounds_simple}
  \Delta V \leq E \leq n\Delta V
\end{align}
for the approximated dilution index $E$. 
In the case of an unbounded domain, the upper limit cannot be reached, while for a bounded domain $\Omega$, the maximal dilution index $E_{\text{max}} = n \Delta V = \text{Area}(\Omega)$ is given as the size of the domain. For a bounded domain, it is convenient to normalize the dilution index by its maximum, yielding the reactor ratio \cite{kitanidis_1994}
\begin{align*}
  M = \frac{E}{E_{\text{max}}}.
\end{align*}

Considering $m$ realizations, we can see each realization as one particle that evolves with the flow.
For a grid with more cells than particles, i.e., $n > m$, we can find a second upper bound to $E$, as maximal $m$ out of the $n$ cells can contain a particle, while all others must be empty. This results in 
\begin{align}\label{eq:E_bounds}
  E \leq m \Delta V < n \Delta V = E_{\text{max}}, \qquad\text{if } m < n.
\end{align}
However, using more grid cells than particles only makes sense if large parts of the domain are particle-free.
It is important to notice that inequality \eqref{eq:E_bounds} is only defined for the approximated dilution index $E$, which depends on the number of particles $m$ and a grid size $\Delta V$. Considering $\Delta V \rightarrow 0$ and keeping $m$ fixed, we yield $E \rightarrow 0$. On the contrary, the approximation of $\mathcal{E}$ in \eqref{eq:E_approx} using the exact values of $\rho_k$ will converge to $\mathcal{E}$ as $\Delta V \rightarrow 0$. Yielding $E \rightarrow \mathcal{E}$ is only possible for $\Delta V \rightarrow 0$ simultaneous to $m \rightarrow \infty$.
Given that $m$ is fixed, selecting a suitable grid size can be compared to the problem of determining an appropriate size of the representative elementary volume (REV) to get the volumetric porosity of a porous medium \cite{bear_1988}. The volume of the REV shall be much smaller than the size of the domain to ensure a low approximation error in \eqref{eq:E_approx} and large enough to include a sufficient amount of particles, ensuring a low error in the approximation of $\rho_k$.

\section{Methodology}\label{sec:methodology}

\begin{figure}[tbp!]
  \centering
  \includegraphics[width=8.6cm]{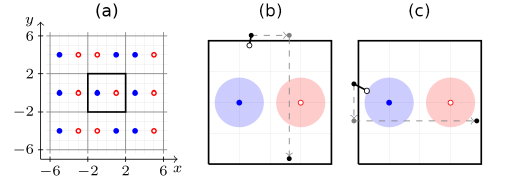}
  \caption{(a) Configuration of sources and sinks that motivates the periodic-like boundary condition which was applied for the PSS system. (b) and (c) show two examples of particle paths that start at the white filled circle and that cross the domain quasi-periodic boundary at the top and at the left side. The blue filled circles indicate the sources, red circles indicate the sinks.}
  \label{fig:PSS_boundary}
\end{figure}

\begin{figure*}[tbp!]
  \centering
  \includegraphics[width=15.2cm]{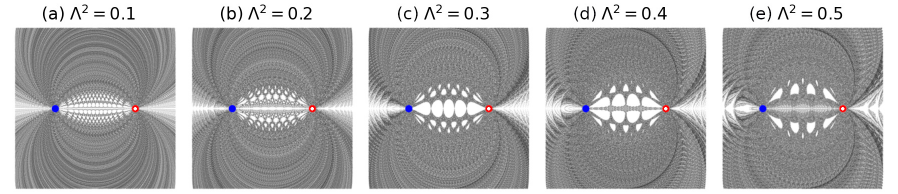}
  \caption{Poincaré sections of the pulsed source-sink system for $\Lambda^2 \in \{0.1, 0.2, 0.3, 0.4, 0.5\}$. The blue filled circle indicates the location of the source while the red circle indicates the sink. White areas not occupied by particles are the KAM islands.}
  \label{fig:PSS_poincare}
\end{figure*}

\subsection{Numerical implementation of the pulsed source-sink system}

As motivated in \cref{sec:PSS_math}, sink and source can be placed at $\pm\textmatrix{1\\0}$ without loss of generality. However, the pulsed source-sink system is unbounded by construction, which is impractical for our application, as this would imply the necessity of an infinite amount of particles. We, therefore, define a bounded domain $[-2, 2]^2$ with periodic-like boundary conditions. Similar to the standard periodic boundary condition, the periodic-like boundary condition can be motivated by a configuration where sources and sinks are repeatedly placed in the $\R^2$ space. Instead of simply repeating the configuration, source and sink are flipped above and below the domain. We show the resulting configuration of sources and sinks in \cref{fig:PSS_boundary}(a). For a particle leaving the domain at the top or bottom, we specify
\begin{align*}
  x_{\text{new}} = -x, \qquad y_{\text{new}} = \begin{cases}
    y-4 & y>0\\
    y+4 & y<0
  \end{cases}.
\end{align*}
Accordingly, for a particle leaving at the left or right
\begin{align*}
  x_{\text{new}} = \begin{cases}
    x-4 & x>0\\
    x+4 & x<0
  \end{cases}, \qquad y_{\text{new}} = -y.
\end{align*}
Two example trajectories of a particle crossing the domain at the top and at the left side of the domain are shown in \cref{fig:PSS_boundary}(b) and (c), respectively.
The modification of the periodic boundary condition prevents an unwanted feature, as we will explain in the following. 
We define the region close to the domain boundary such that all particles that stop inside this region after the sink operation will be pushed out of the domain during the next operation of the source. This definition is reasonable, as particles can only cross the domain boundary during the operation of the source. We also remark that particles will move further due to the operation of the source or sink the closer they are to the source or sink, respectively. Hence, a particle with location $x < 0$ will always move further during the source operation than during the sink operation. We consider a particle with $x < 0$ to stop inside the boundary region at the top or bottom of the domain after the operation of the sink. Due to the definition of the boundary region, the particle crosses the domain boundary during the next operation of the source.
If this particle reenters the domain with the standard periodic boundary condition, it stops again in the boundary region on the negative side of the $x$-axis. For $x < 0$, the sink is too weak to pull this particle out of this region. Hence, it will again be pushed back out of the domain during the next stroke of the source. Using the periodic-like boundary condition, the particle crossing the domain boundary with $x<0$ will always reenter with $x>0$ (\cref{fig:PSS_boundary}(b) and (c)). For $x>0$, the sink is strong enough to pull the particle out of the boundary region before the next operation of the source. Hence, using the periodic-like boundary condition instead of the standard one, we prevent particle trajectories alternating between the upper and lower boundary with each source operation.
Upon this first unwanted feature, the standard periodic boundary condition also results in a big particle-free region on the positive side of the $x$-axis. On this side, the sink is stronger than the source; therefore, the particles will be extracted from the domain without reaching certain areas. With our periodic-like boundary condition, particles crossing the domain boundary with $x<0$ will reenter with $x>0$ (\cref{fig:PSS_boundary}(b) and (c)); as they approach the sink radially, they reach the parts of the domain on the positive side of the $x$-axis that would otherwise stay particle-free.

Benchmarking the evaluation time of $\Psi_\sigma$ for the pulsed source-sink system when tracking a particle for time $\Delta t$ on an Intel i5 processor yields about $0.02\,$ms. This runtime restricts the number of particles used in the particle tracking experiments. Choosing a sufficient amount of particles is crucial to achieving a high resolution of the particle density approximation \cite{schmidt_2017}, especially for mixing processes limited by diffusion \cite{rahbaralam_2015}, as in our case. The number of particles $m$ limits the resolution of low-concentration regions as a grid cell can only contain an integer number of particles. Hence, the particle density of a non-empty cell $V_k$ can not be smaller than $\frac{1}{m\,\Delta V_k}$. Following the argumentation of \cite{rahbaralam_2015}, the simulation of each solute molecule in the system is impossible due to their enormous number. Therefore, we restrict ourselves to $m = 10^7$ particles, which is inside their proposed range of $10^6$--$10^9$.

To parametrize the system, we have one parameter $\Lambda \in (0, \infty)$. Its definition in \eqref{eq:Lambda} requires $a$, $Q$ and $\tau$. Thereby, different combinations may result in the same behavior of the system. We hence fix $a=1$ and $Q=4\pi$ for all experiments. This yields $\Lambda = \sqrt{4\tau}$, which is modified by different choices of $\tau$. 
The choice of $Q$ is motivated by doubling the areal discharge implicitly used for source and sink in the RPM flow.
Given $t_{\text{max}} = 20$ in our experiments, we can determine the number of strokes of source and sink by $t_{\text{max}}/(2 \tau) = 10/\tau$.
Given that all particles with distance $< \Lambda$ to the sink are extracted during the next stroke of the sink, we need to ensure that $\Lambda$ is sufficiently small such that the circles depicted in \cref{tikz:exp_setup} have some distance to the domain boundary. We hence restrict $\Lambda^2$ to the interval $(0, 0.5]$. The final selection for the experiment is $\Lambda^2 \in \{0.1, 0.2, 0.3, 0.4, 0.5\}$. For completeness, \cref{fig:PSS_poincare} shows Poincaré sections for these parameters. Poincaré sections show the locations of particles tracked after each period of operation. In our case, these plots were produced by tracking five particles that started inside the chaotic region of the domain and saving their location after each time $2\tau$. Tracking a particle corresponds to repeatedly applying the Poincaré map \eqref{eq:PSS_M}. 
We observe particle-free island inside the plots, which are the KAM islands. These islands consist of periodic points with various periods \cite{aref_2017}. They originate from the neighborhood of elliptic points and are invariant under the advective flow $\Psi$ \cite{ottino_1989,lester_2009,chate_1999}. Due to the periodic nature of the islands, particles that start inside the chaotic region of the domain cannot enter these islands. Therefore, these islands stay free of particles in the Poincaré sections.

\subsection{Numerical implementation of the RPM flow}

\begin{figure*}[tbp!]
  \centering
  \includegraphics[width=15.2cm]{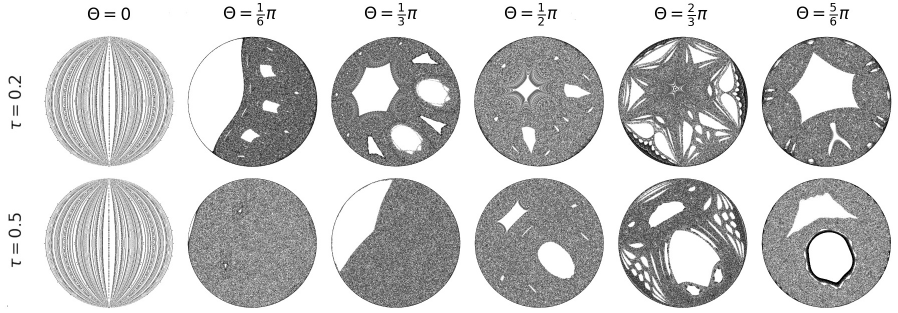}
  \caption{Poincaré sections of the RPM flow for $\Theta \in \left\{0, \pi/6, \pi/3, \pi/2, 2\pi/3, 5\pi/6\right\}$ and $\tau \in \{0.2, 0.5\}$. White areas not occupied by particles are the KAM islands.}
  \label{fig:RPM_poincare}
\end{figure*}

In this setup, the boundary condition is less critical than in the previous case. Unlike the PPS system, where particles leave the domain due to the strokes of the source, particles in the RPM flow can leave the system only due to diffusion. Thus, introducing standard reflective boundary for our domain is sufficient.  When computing the advective particle movement in the RPM flow, solving equation \eqref{eq:RPM_formular_advection} is challenging. Since \eqref{eq:RPM_t_adv} cannot be solved for $\theta$ in terms of $(t_{\text{adv}}, \psi)$, we perform a numerical inversion over a high-resolution grid, as proposed in \cite{lester_2009}. The results in this paper use an equidistant grid with $1,000 \times 1,000$ nodes in the $\{\theta, \psi\}$ space and a piecewise linear barycentric interpolation to approximate the function $t_{\text{adv}}^{-1}(t_{\text{adv}}, \psi)$. To verify this approach, we computed the error approximation $|\theta - (t_{\text{adv}}^{-1} \circ t_{\text{adv}})(\theta, \psi)|$ using $10,000$ uniformly distributed random samples in the right half sphere, yielding a mean error in $\mathcal{O}(10^{-5})$. The same analysis was also performed with $100 \times 100$ and $10 \times 10$ nodes, yielding a mean error in $\mathcal{O}(10^{-3})$ and $\mathcal{O}(10^{-1})$, indicating that the approximation error decreases linearly with the number of nodes. 
In addition to an initialization time of about $30\,$s to approximate $t_{\text{adv}}^{-1}$ on an Intel i5 processor, an evaluation of $\Psi_\sigma$ for the RPM flow takes about $0.2\,$ms, ten times slower than for the pulsed source-sink system. To balance computational accuracy and runtime, we decided to reduce the number of particles in our numerical experiments to one million. The results are less accurate than those of the pulsed source-sink system, but the numerical errors are sufficiently small to draw conclusions.

To control the system, we set the rotation angle $\Theta$ and the time $\tau$ between rotations. As shown in \cite{lester_2009}, the parameter space is restricted to $\{\Theta, \tau\} \in [0, \pi] \times [0, \infty)$. Setting formally $\tau < 0$ results in switching the position of the source and the sink, which is equivalent to a rotation of the domain by $180$ degree. In general, we can restrict $\Theta \in [-\pi, \pi]$. Additionally, the reflection symmetry of the base flow shows the equivalence of $-\Theta$ to $\Theta$, which gives the desired result. Alike for the pulsed source-sink system, we use $t_{\text{max}} = 20$, restricting the $\tau$ parameter space further to $(0, t_{\text{max}}]$. For the experiments, we use the Cartesian product over the two sets of parameters $\Theta \in \left\{0, \pi/6, \pi/3, \pi/2, 2\pi/3, 5\pi/6\right\}$ and $\tau \in \{0.2, 0.5\}$. The Poincaré sections in \cref{fig:RPM_poincare} were created by repeatedly applying the Poincare map \eqref{eq:RPM_M} on three particles that started in the chaotic region. We again see KAM-islands filling large parts of the domain. Compared to the pulsed source-sink system, the percentage of the domain covered by islands is bigger, while the absolute size might be equal or even smaller due to the difference in the domain size.
We also observe more differences between the results for each parameter configuration. 
For $\Theta = 0$, the source and sink do not rotate, and the particles always stay on the trajectory in the reference system they are initially placed on. In this configuration, no chaotic flow structures can be observed. For $\Theta > 0$, we observe KAM islands. The size of the region covered by islands differs for each parameter configuration. As we observe chaotic advection only between the KAM islands in the chaotic region, the KAM island structure particularly impacts the chaotic mixing of the flow. We expect increased mixing for $\Theta = \pi/6$ and $\tau = 0.5$, where we only find three very small islands. On the contrary, for $\Theta = \pi/6$ and $\tau = 0.2$, nearly half of the domain is particle-free in \cref{fig:RPM_poincare}, indicating less mixing enhancement by chaotic advection. 

\begin{figure}[tbp!]
  \centering
  \includegraphics[width=8.6cm]{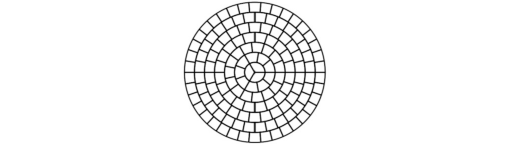}
  \caption{Grid with $h = 0.15$ for approximating the dilution index of the RPM flow.}
  \label{fig:RPM_grid}
\end{figure}

A final challenge for the RPM flow is the circular domain. In order to compute the dilution index, we need to construct a grid. Because of the circular shape, a rectangular or triangular grid is insufficient due to the intersection of grid cells with the domain boundary. Instead, we create a grid in the space of polar coordinates where we have levels of width $h$ for the radius and choose a suitable step size for the angle to yield $\Delta V \approx h^2$. This ensures that the cells have comparable diameters. We give an example grid with $h = 0.15$ in \cref{fig:RPM_grid}. Grid sizes may vary slightly over the different levels, therefore, we use the formula for the dilution index over nonuniform grids in \eqref{eq:E_nonuniform}.

\subsection{Time stepping}

The flow defined in \eqref{eq:diffusive_basic_flow} consists of an advective and a diffusive part. While the advective part is solved analytically, the diffusive part is added after a discrete time step $\Delta t = 0.01$. We introduce a non-diffusion region around the source and sink to prevent numerical errors due to the time discretization. This region prevents particles that stop close to the sink or source from being affected by diffusion. In this region, the particle trajectory is very sensitive to small perturbations; due to diffusion, a particle would significantly change its angle towards the source or sink, which is unrealistic. Since particles get infinitely close to the sink or source, the problem remains even when reducing the time step.
From an experimental point of view, these regions mimic the finite-dimensional circular inlet/outlet of the sink/source. Particles that reach the sink within time $0.01$ are assumed to be already in the pipe between the source and the sink, where we assume that no diffusion occurs. The same applies to particles that have been injected by the source within time $0.01$.

\subsection{Diffusion coefficient}

In order to compare the PPS and the RPM systems, we define a relationship between the diffusion coefficient $\sigma$ based on the Péclet number. Therefore, the results of the systems are comparable, independent of the different domain sizes and flow velocities. The Péclet number is defined as $\Pe = Lu/\sigma^2$, with characteristic length $L$, flow velocity $u$ and diffusion coefficient $\sigma$. For the pulsed source-sink system, we alternate between $v_-$ for operating the sink and $v_+$ for operating the source. Using $L_{\text{PSS}}=4$ and
\begin{align*}
  u_{\text{PSS}} = \int_{[-2, 2]^2} \left|\left|v_-\right|\right|\, d(x,y) = \int_{[-2, 2]^2} \left|\left|v_+\right|\right|\, d(x,y) \approx \frac{Q}{2\pi} 13.342
\end{align*}
yields $\Pe_{\text{PSS}} \approx 53.368\frac{Q}{\pi\sigma^2}$. Analog for the RPM flow 
\begin{align*}
  u_{\text{RPM}} = \int_{B_1\textmatrix{0\\0}} \left|\left|v_{\text{RPM}}\right|\right| \, d(x,y) \approx 7.328
\end{align*}
and $L_{\text{RPM}} = \sqrt{\pi}$ yield $\Pe_{\text{RPM}} \approx 12.989/\sigma^2$.
To match the Péclet numbers of both systems, we define a relationship between the diffusion coefficient $\sigma$, yielding
\begin{align*}
  \sigma_{\text{PSS}}
  \approx 2.027 \sqrt{\frac{Q}{2\pi}}\, \sigma_{\text{RPM}} =: \beta \,\sigma_{\text{RPM}}.
\end{align*}
For the experiments, we define two different levels of diffusion $\sigma_{\text{RPM}}\in\{0.01, 0.1\}$ and $\sigma_{\text{PSS}}\in\{0.01\beta, 0.1\beta\}$. The final Péclet numbers lie in $\mathcal{O}(10^{5})$ and $\mathcal{O}(10^{3})$ showing that the setups are clearly advection dominated.

\subsection{A computationally reliable approach for the approximation of the dilution index}\label{sec:dil_gaussian}

The definition of the dilution index uses a zero-order estimation method to approximate the particle density in the domain, called nearest grid point method \cite{peterka_2016}. Decreasing the grid size leads to approximation errors in the estimation of the mean probability density $\rho_k \approx \frac{P_k}{\Delta V_k}$ inside each cell $k = 1, ..., n$.
For a given number of particles $m$ and an equidistant grid of grid size $\Delta V$, we have derived an upper bound for the dilution index $E$ in \eqref{eq:E_bounds}. Decreasing $\Delta V \rightarrow 0$ results in $E \leq m \Delta V \rightarrow 0$ independent of the underlying particle distribution. An unrealistic result. Given the distribution of $m$ particles, our goal is to establish a sophisticated approach to select a grid size sufficiently small to reflect fine structures in the plume but big enough to ensure a low error in the approximation of $\rho_k$.
The obtained density approximation is called "fine-grained density" \cite{krasnopolskaya_2009}. 

We assume an instantaneous pulsed injection. To select an appropriate grid size, we compare the numerical approximation of the dilution index against its analytical solution. Given the diffusion coefficient $\sigma$, the plume is Gaussian distributed with variance $\sigma^2 t$ at time $t$ \cite{kitanidis_1994}. We can compute the analytical dilution index by inserting the corresponding Gaussian density function $\rho(x, t)$ into \eqref{eq:E_continous}, yielding
\begin{align}\label{eq:E_analtical}
  \begin{split}
  \mathcal{E}(t) = \exp\left( -\int_\Omega \rho(x, t) \ln(\rho(x, t)) \, dx \right) 
  = \exp(\ln(2\pi\sigma^2t)+1)
  = 2 \pi \sigma^2 t \exp(1).
  \end{split}
\end{align}

\Cref{fig:gaussian} shows the results for $t=1$ for different diffusion coefficients and $m = 10\,000$ particles. When $\sigma=0$ (black color), one cell contains all particles and $E = \Delta V = h^2$; thus $E$ decreases quadratically with $h$ (\cref{fig:gaussian}(a)). As shown in \eqref{eq:E_bounds_simple}, this is a general lower bound for $E$. When $\sigma > 0$, for big grid sizes, the dilution index converges to the analytical dilution index (\cref{fig:gaussian}(b)). By decreasing the grid size, we reach an optimal grid size $h^*$ with error zero. Notice that \cref{fig:gaussian} was created using only a finite selection of grid sizes that do not include $h^*$. Thus, the approximation error is always greater than zero. For smaller grid sizes, the error increases again due to increasing errors in the approximations of $\rho_k$, and the dilution index approaches the upper bound $m \Delta V$ as $h$ tends to $0$ (\cref{fig:gaussian}(a)). 
This results in a plateau in the plot $E$ over $h$ at the value of the analytical dilution index. The length of the plateau varies depending on the number of particles used in the experiments. In general, at the center of the plateau, it holds that $\frac{dE}{dh}(h^*) = 0$. For a discrete selection of grid sizes, the one closest to $h^*$ may be found by estimating the minimum of the forward finite differences. 
This approach has the problem that with $h \rightarrow 0$, we yield $E \rightarrow 0$ quadratically and hence also $\frac{dE}{dh}(h)\rightarrow 0$. Hence, finding the minimal derivative of $E$ using forward finite differences over a discrete grid may select erroneously small grid sizes. Instead, we use the derivative of $\ln(E)$ with respect to the logarithmic scaled grid size $h$. This alternative term is also zero if and only if $h = h^*$. We compute the derivative using forward finite differences with step size $\delta > 0$
\begin{align}\label{eq:loglog_derivative}
  \Delta_{\delta} [E](h) := \frac{\ln(E(h\exp(\delta))) - \ln(E(h))}{\delta}.
\end{align}
Multiplying $h$ by $\exp(\delta)$ is equivalent to taking a $\delta$ step from $h$, respecting the logarithmic scale. This approach has the advantage that $\Delta_{\delta} [E](h) \rightarrow 2$ as $h \rightarrow 0$. Comparing the derivative in \cref{fig:gaussian}(c) to the relative approximation error in \cref{fig:gaussian}(b), we see that the grid sizes of low approximation errors indeed lie at the center of the plateau. This observation motivates the choice of the grid size $h$ to be the minimum of the derivative. Independent of the used diffusion coefficient $\sigma$, we observe similar relative errors close to the optimal grid size $h^*$. The more particles used, the clearer the minimum.

This approach also aligns with the theory of REV used to compute a homogenization of the volumetric porosity of a porous medium. Lowering the size of the REV below a particular value $\Delta V_0$ results in large fluctuations in the ratio of void space in the representative volume compared to its size. As the size goes to zero, the ratio takes a value in $\{0, 1\}$, depending on the centroid of the REV. Equal to 0 if the centroid is inside the pore, or equal to 1 if it is inside the solid matrix. This leads to an undesirable result of the homogenization. To determine the volumetric porosity, the limit $\Delta V_i \rightarrow \Delta V_0$ is used \cite{bear_1988}. 

\begin{figure}[tbp!]
  \centering
  \includegraphics[width=8.6cm]{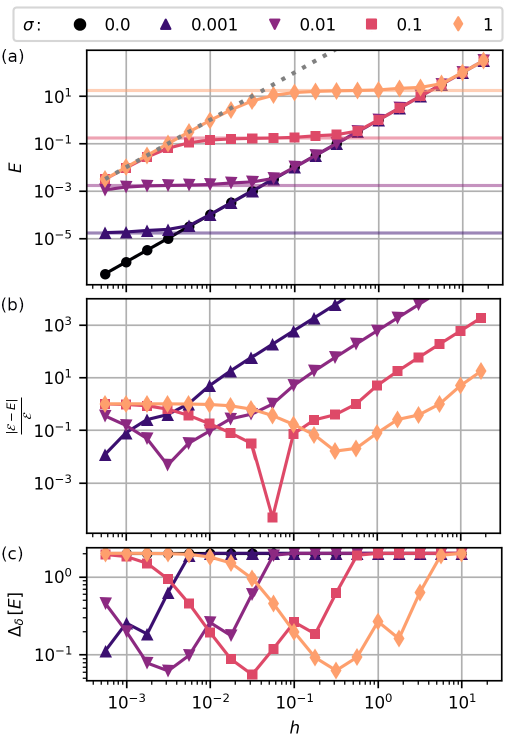}
  \caption{(a) Shows the value of the dilution index $E$ over the grid size $h$ for $m = 10\,000$ particles and different diffusion coefficients $\sigma$. The dotted line gives the upper bound of the dilution index $m\Delta V$. 
  The continuous lightened lines help to visualize 
  the plateaus forming at the value of the analytical dilution index.
  (b) Shows the relative error $\frac{|\mathcal{E} - E|}{\mathcal{E}}$ of the numerical dilution index $E$ depending on the grid size $h$ to the analytical value $\mathcal{E}$ from \eqref{eq:E_analtical}. We observe a convergence $E \rightarrow \mathcal{E}$, until we pass the optimal grid size $h^*$. For smaller $h$, the dilution index dissipates from the analytical value.
  (c) Shows the approximated derivative of the dilution index $\Delta_\delta[E]$, using forward finite differences in the log-log scale as defined in \eqref{eq:loglog_derivative}, for $\sigma \neq 0$. The minima of the derivatives correspond approximately to the minima in the approximation error of plot (b).}
  \label{fig:gaussian}
\end{figure}

The example of the pulse injection also demonstrates the importance of a sophisticated way of choosing a suitable grid size. Fixing one grid size for all diffusion coefficients leads to huge errors in some dilution index approximations. We also find the number of particles $m$ to be crucial. Using more particles increases the accuracy of the approximation of $\rho_k$. This allows for a smaller grid size $h^*$, resulting in a higher accuracy in the approximation. A more detailed figure that further supports our findings concerning the number of particles $m$ is included in the supplementary material section S2 Fig. S1.

\begin{figure*}[tbp!]
  \centering
  \includegraphics[width=15.2cm]{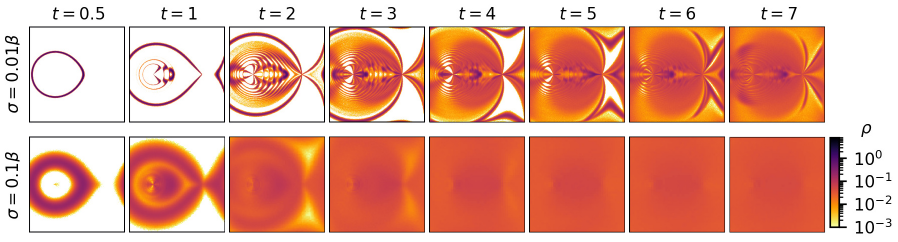}
  \caption{Evolution of the probability density of the particle distribution $\rho$ over time using $\Lambda^2 = 0.2$ and $\sigma \in \{0.01\beta, 0.1\beta\}$ for the pulsed source-sink system.}
  \label{fig:PSS_density_evolution}
\end{figure*}

\begin{figure}[tbp!]
  \centering  
  \includegraphics[width=8.6cm]{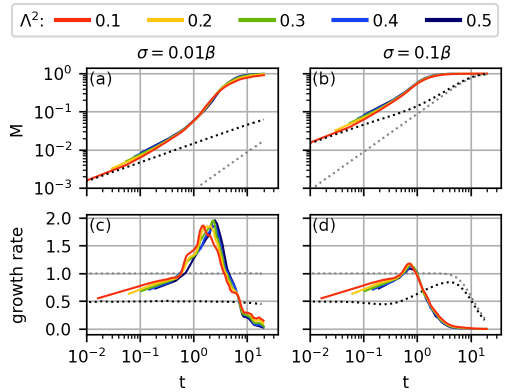}
  \caption{(a)-(b) Reactor ratio $M$ and (c)-(d) reactor ratio growth rate over time $t$ for the PSS system with different values of $\Lambda$. The growth rate was approximated using the moving average over the first order finite differences with a window of $\pm 0.05$ in the exponent. The dotted lines correspond to the case without advection where the initial particle position is either at one point (gray) or in a ring (black).}
  \label{fig:PSS_growth}
\end{figure}

\begin{figure}[tbp!]
  \centering
  \includegraphics[width=8.6cm]{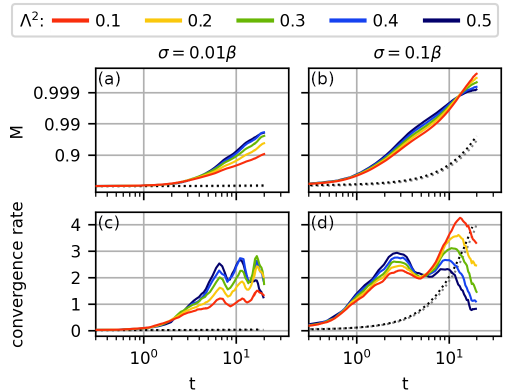}
  \caption{(a)-(b) Reactor ratio $M$ plotted with a logarithmically typed $y$-axis highlighting the convergence $M \rightarrow 1$ and (c)-(d) convergence rate of the reactor ratio for different values of $\Lambda$ in the PSS system computed using \eqref{eq:convergence_rate}. The convergence rate was approximated with the moving average over the first order finite differences with a window of $\pm 0.05$ in the exponent. The dotted lines correspond to the case without advection where the initial particle position is either at one point (gray) or in a ring (black). Note that for $\sigma = 0.01\beta$ the difference between the results of point and the ring initialization is below the resolution of the plot.}
  \label{fig:PSS_convergence}
\end{figure}

\section{Results}\label{sec:num_experiments}

\subsection{Pulsed source-sink}

For the experiments with the pulsed source-sink system, we used $10$ million particles, a time step $\Delta t = 0.01$ as well as the flow parameters motivated in the previous section $\Lambda^2 \in \{0.1, 0.2, 0.3, 0.4, 0.5,\}$, $\sigma \in \{0.01\beta, 0.1\beta\}$ and $t_{\text{max}} = 20$. 
The initial position of the particles is at the location of the source. The source operates first and after time $\tau$ the particles form a circle of radius $\Lambda$ around $\textmatrix{-1\\0}$. The reactor ratio was approximated only at the end of the strokes of the source when all extracted particles have been reinjected, at times $t=(2n + 1)\tau$ with $n \in \N$. This ensures that the dilution index is estimated over a mass conservative plume. For comparison, we perform two baseline experiments without advection. First, we use a point initialization for which all particles are initially placed at the origin. 
In the second baseline experiment, we take the locations of the particles of the advection experiment with $\Lambda^2=0.1$ at time $t = 0.05$ as initialization.
To ensure consistency with the previous experiments we set the initial time to $t = 0.05$. At early times the particles form a ring around the source, and thus we refer to this initial setup as ring initialization.
For each considered time step we select a grid size $h$ for the dilution index approximation using the minimum of the finite differences \eqref{eq:loglog_derivative}. During our experiments, we often find several local minima. Due to the discrete choice of grid sizes an observed local minimizer may actually be closer to $h^*$ than the observed global minimizer. Given that the dilution index only grows in time, we choose the smallest local minimum to ensure that $h$ increases monotonically in time. To avoid very big grid sizes we set $h \leq 0.1$ leading to a minimum of $1600$ grid cells in the domain. The estimated grid size is also ideal for visualizing the particle density.

We expect the reactor ratio to increase monotonically over time while approaching $M_{\text{max}} = 1$. We also expect faster dilution in the chaotic advection setups compared to the baseline experiments. In \cref{fig:PSS_density_evolution}, it is possible to see the effect of chaotic advection on the evolution of the probability density of the particle distribution for $\Lambda^2 = 0.2$. For this and all other considered flow parametrization, we show the approximated reactor ratio $M$ over time $t$ in \cref{fig:PSS_growth}(a) and (b). \cref{fig:PSS_growth}(c) and (d) show the reactor ratio growth rate, which is always positive, demonstrating that our method preserves the monotonical increase.
The baseline cases without advection display constant growth rates for small times.
For the point initialization, the plume is distributed according to the Gaussian distribution with variance $\sigma^2 t$ for small $t$. The reactor ratio is therefore given by $M(t) = \pi/8 \exp(1) \sigma^2 t$, which can only increase linearly in time. 
The ring initialization grows at a constant rate of $0.5$ for small times. The reason is that diffusion can only act in the direction perpendicular to the ring due to the ring structure. As shown in \cite{kitanidis_1994}, the dilution index of the Gaussian distribution in one dimension with variance $\sigma^2 t$ is given by $\sqrt{2\pi t}\, \sigma^2 \exp\left(1/2\right)$ which has a growth rate of $0.5$ in time. For later times, the particles fill the ring and appear more similar to a two-dimensional Gaussian distribution. Therefore, the growth rate can reach up to one, and for later times and $\sigma = 0.1\beta$, the reactor ratios of both baseline experiments coincide approximately.
Despite the lower growth rate, we note that the reactor ratio for the baseline with ring initialization is higher than for the point initialization. This behavior is expected due to the higher entropy of the particle location.
In the chaotic advection systems, the growth rates for small times are also close to $0.5$. At the beginning, the particles form a circle around the source, and thus, the behavior is similar to the baseline with the ring initialization. As the particles move away from the source, the growth rate increases slowly until $t \approx 0.6$ when the growth rate experiences a remarkable rise due to the first cluster of particles reaching the sink. As we chose $Q$ to be the same for all configurations, the particles arrive at the sink within a similar time frame, explaining the simultaneous rise of the growth rates for all experiments, i.e., for different choices of $\Lambda^2$ and $\sigma$. Due to the boundedness of the reactor ratio, the growth rate cannot remain at a high level for a long time. The rate quickly falls to zero, especially for $\sigma = 0.1\beta$ (\cref{fig:PSS_growth}(d)). As the particles fill the domain in \cref{fig:PSS_density_evolution} bottom $M$ approaches $M_{\text{max}}$. However, this already shows the increased dilution caused by chaotic advection.
Considering now \cref{fig:PSS_convergence}(a) and (b), which also show the graph $M$ over $t$, but the axis for the reactor ratios have a logarithmic scaling to highlight how $M$ approaches $1$. For larger times and $\sigma = 0.1 \beta$, the reactor ratio reaches a maximum degree of dilution, which is strictly smaller than $1$. This effect is due to low particle density regions between domain boundary and source/sink also observed in the Poincaré sections in \cref{fig:PSS_poincare}, which is an artifact of the periodic-like boundary condition. Looking at the convergence rates of the reactor ratio, approaching $M_{\text{max}}$ in \cref{fig:PSS_convergence}(c) and (d) and represent the visible derivative of the curves in \cref{fig:PSS_convergence}(a) and (b). The convergence rates at time $t$ can be computed numerically via 
\begin{align}\label{eq:convergence_rate}
  \frac{\ln(1 - M(t)) - \ln(1 - M(t + \Delta t))}{\ln(t+\Delta t) - \ln(t)}.
\end{align}
We observe a short-time increase in the convergence rate each time a larger cluster of particles approaches a less occupied region of the domain. In particular, the cluster responsible for the bumps in the convergence rate originates from particles injected by the source onto trajectories, which lead them approximately straight upwards or downwards towards the domain boundary. These trajectories are the slowest regarding travel distance from source to sink. The cluster hits the sink at $t \approx 6$, as seen in the top row of \cref{fig:PSS_density_evolution}, filling the particle-free region above and below the sink.

\subsection{RPM flow}

\begin{figure}[tbp!]
  \centering
  \includegraphics[width=8.6cm]{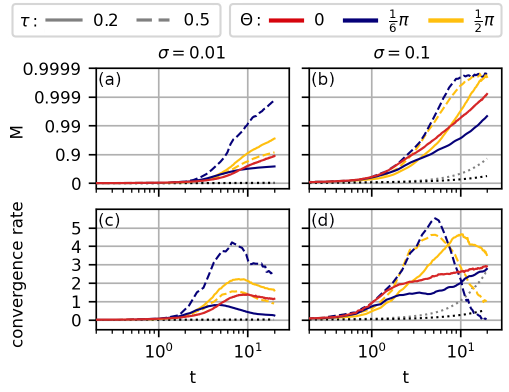}

  \caption{(a)-(b) Evolution of $M$ over $t$ for the RPM flow with a logarithmically typed $y$-axis and (c)-(d) the convergence rate for $M$ approaching $1$ for different flow parametrization computed using \eqref{eq:convergence_rate}. The convergence rate was computed using the moving average over the first order finite difference by using a window of $\pm 0.2$ in the exponent. The dotted lines correspond to the case without advection where the initial particle position is either at one point (gray) or in a ring (black). Note that for $\sigma = 0.01$ the difference between the results of point and the ring initialization is below the resolution of the plot.}
  \label{fig:RPM_convergence}
\end{figure}

\begin{figure*}[tbp!]
  \centering
  \includegraphics[width=17.8cm]{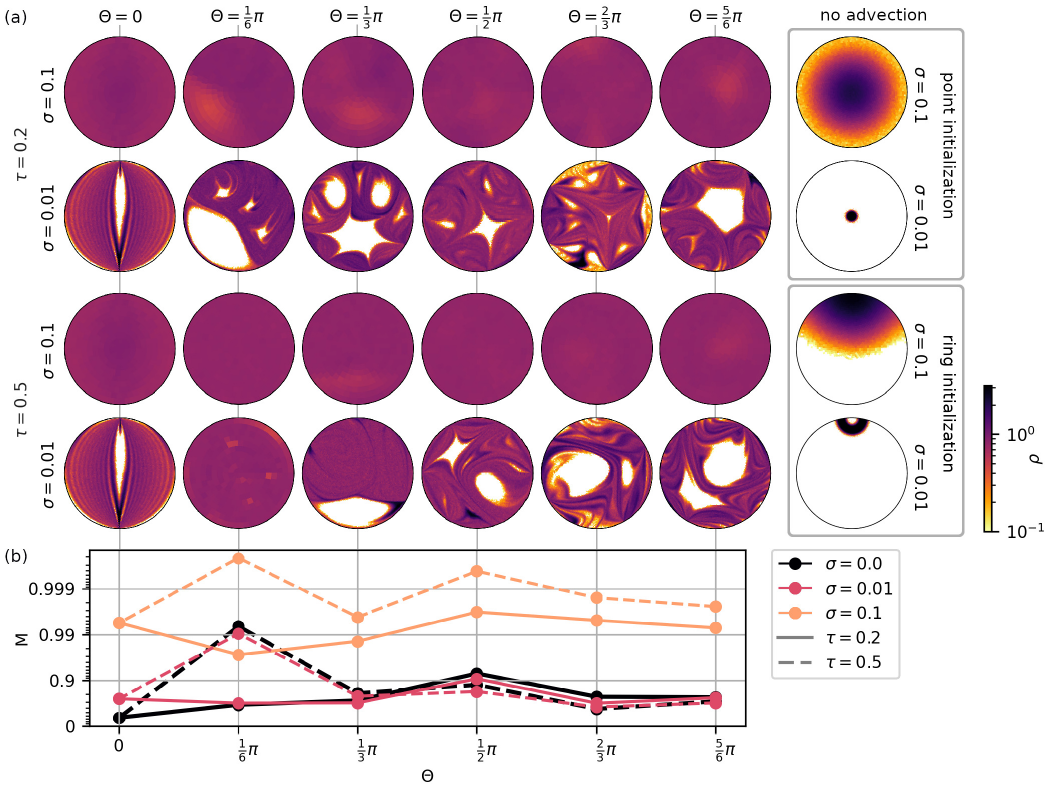}
  \caption{(a) Probability density approximation at time $t = 10$, and (b) reactor ratio $M$ over $\Theta$ for different $\Theta$, $\tau$ and $\sigma$ for the RPM flow.}
  \label{fig:RPM_dil_20}
\end{figure*}

During the RPM system experiment, $1$ million particles are initially placed at the location of the source. This initialization represents that they are injected through the source at $t=0$. The parameters are chosen as motivated above with the rotation angle $\Theta \in \left\{0, \pi/6, \pi/3, \pi/2, 2\pi/3, 5\pi/6\right\}$, the time between rotations $\tau \in \{0.2, 0.5\}$, the diffusion coefficient $\sigma \in \{0.01, 0.1\}$, $t_{\text{max}}= 20$ and $\Delta t = 0.01$. While $\Theta > 0$ displays chaotic advection, we do not find chaotic flow structures for $\Theta = 0$. However, we observe shear flow and focusing and defocusing flow, which also leads to deformation of the plume and, therefore, to increased mixing \cite{rolle_2019,paster_2015}. Like for the pulsed source-sink system, we consider two baseline setups without advection, but with the same reflective boundary conditions as for the systems with advection. We also use the point initialization, where all particles are initially placed at the origin, and the ring initialization, where we take the locations of the particles of the advection experiments at time $t = 0.01$ as initialization. When approximating the dilution index, we estimate the grid size $h$ for each considered time step individually, by using the minimum of $\Delta_\delta[E](h)$ defined in \eqref{eq:loglog_derivative}, ensuring that $h$ is monotonically increasing in time and $h \leq 0.1$. 

\begin{figure*}[tbp!]
  \centering
  \includegraphics[width=15.2cm]{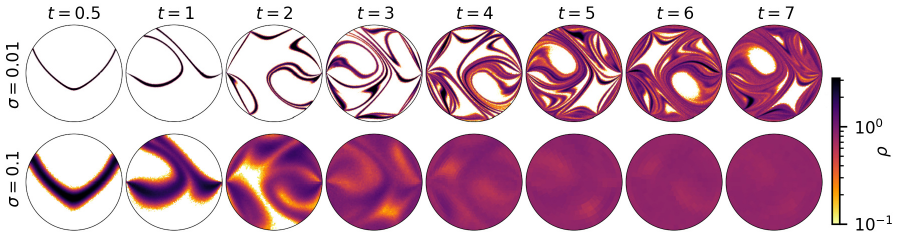}

  \caption{Evolution of the probability density of the particle distribution over time using $(\Theta, \tau) = \left(\pi/2, 0.5\right)$ and $\sigma \in \{0.01, 0.1\}$ for the RPM flow.}
  \label{fig:RPM_density_evolution}
\end{figure*}

In \cref{fig:RPM_convergence}(a) and (b) we show the evolution of the reactor ratio $M$ over $t$ for $\Theta \in \left\{0, \pi/6, \pi/2\right\}$ and $\tau \in \{0.2, 0.5\}$. Fig. S2 in section S3 of the supplementary material includes the results for all considered parameter combination. The $y$-axis has a logarithmic typed rescaling to highlight the convergence $M \rightarrow 1$. For smaller times, the reactor ratio increases monotonically, as expected. For $\sigma = 0.1$ and large times, some configurations reach a status of full dilution. Due to the finite amount of particles, the approximated density distribution always differs slightly from the uniform distribution; therefore, the reactor ratio must be smaller than $1$, and we observe oscillations resulting in occasional decreases in $M$. Comparing the dilution enhancement of different flow parameter configurations, we notice significant differences.  
Investigating this further, \cref{fig:RPM_dil_20}(a) and \cref{fig:RPM_dil_20}(b) show the probability density approximations and $M$ over $\Theta$, respectively, at time $t = 10$, where reactor ratios differ remarkably between the configurations. Each figure column represents a rotation angle $\Theta$, and \cref{fig:RPM_dil_20}(a) includes the cases without advection on the right. For $\sigma = 0.01$ and $\Theta > 0$, the KAM islands found in the Poincaré maps of \cref{fig:RPM_poincare} are mainly particle-free. This is due to the periodic nature of the points inside the islands. As the particles are not initially placed on trajectories leading through these islands, they can only enter due to the diffusion process. This makes the dilution inside these regions far slower than in the chaotic area of the domain. For $\Theta = 0$, no KAM islands arise in the flow, but due to the non-chaotic nature of the flow, particles fill the domain much slower. 
We, therefore, achieve faster dilution for chaotic flow fields with only a few small KAM islands as for $(\Theta, \tau) = \left(\pi/6, 0.5\right)$ or $\left(\pi/2, 0.2\right)$. In general, this implies that the non-chaotic configuration $\Theta = 0$ is not necessarily worse than $\Theta > 0$ in terms of mixing, in particular when compared with configurations leading to big KAM islands.
This effect does not occur in the pulsed source-sink system where all trajectories, including the periodic ones, pass through the sink and the source. By initially placing the particles on the source, some of them end up on the periodic trajectories and fill the KAM islands. 

To quantify the effect of KAM islands, we introduce a quasi reactor ratio. To do so, we simulate the motion of particles with $\sigma = 0$ and approximate the size of the chaotic region. We use a grid with size $h = 0.01$ and count all cells with at least one particle. 
The quasi reactor ratio $\bar{M}(\sigma = 0)$ is given as the fraction of the chaotic area to the full domain (black lines in \cref{fig:RPM_dil_20}(b)). We see that the reactor ratios approximately follow the same trend as the quasi reactor ratio. The parameter configuration $\left(\Theta, \tau\right) = \left(\pi/6, 0.5\right)$ leads to the highest quasi reactor ratio and corresponds to the configuration with the fastest dilution. 

\begin{figure}[tbp!]
  \centering
  \includegraphics[width=8.6cm]{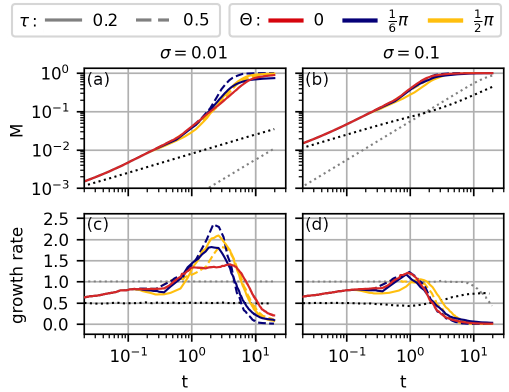}

  \caption{(a)-(b) Reactor ratio $M$ and (c)-(d) reactor ratio growth rate over time $t$ for different flow parametrization for the RPM flow. The growth rate was approximated using the moving average over the first order finite differences with a window of $\pm 0.2$ in the exponent. The dotted lines correspond to the baseline cases without advection where the initial particle position is either at one point (gray) or in a ring (black).}
  \label{fig:RPM_dil_over_t_loglog}
\end{figure}

Summarizing the above, it appears that the individual growth of the reactor ratio is very dependent on the KAM island structure. Therefore, the growth rates and convergence rates of the reactor ratio vary over the parameter space. 
Looking at the convergence rates in \cref{fig:RPM_convergence}(c) (computed using \eqref{eq:convergence_rate}) for $\sigma = 0.01$, we find that they reach a maximum before the rate decreases again. To understand this behavior we look at the temporal evolution of the probability density distributions of the particles evolved using $\Theta = \pi/2$ and $\tau = 0.5$ in the top row of \cref{fig:RPM_density_evolution}. At the start of the time interval of high convergence rates $t \approx 2$, the particles are concentrated at a few stripes in the domain. From here, they quickly spread until the particles are well diluted inside the chaotic area. To also reach high dilution inside the KAM island, particles need to diffuse into the islands.
Since in advective dominated flows diffusive transport is more time-consuming than advective transport the convergence rates of $M \rightarrow 1$ are lower after reaching high dilution in the chaotic area.
For a higher diffusion coefficient (e.g., $\sigma = 0.1$), entering the KAM islands is not a time-consuming process (\cref{fig:RPM_density_evolution}, bottom). For later times, we observe convergence rates close to zero, which is a strong indicator of a setup to be fully diluted.
These results demonstrate that diffusion plays an important role in systems with chaotic advection, as it is the limiting process for filling of the KAM islands.

Finally, we consider the growth rates of $M$ in \cref{fig:RPM_dil_over_t_loglog} and compare the results of the advective setups to the growth rates of the setups without advection. For the results including all considered parameter configurations we refer to the supplementary material Fig S3. As the first rotation happens at time $t = \tau$, the graphs coincide for $t < \tau$, and the growth rate for small times is close to $0.5$ and increases in time.  After some rotations of source and sink, the growth rates may eventually start rising above $1$, especially for $\sigma = 0.01$, where most of the setups reach a maximal growth rate of around $2$ (\cref{fig:RPM_dil_over_t_loglog}(c)). At higher diffusivity, such as $\sigma = 0.1$ (\cref{fig:RPM_dil_over_t_loglog}(d)), the reactor ratio gets close to the maximum much earlier, resulting in lower maximal growth rates. Unlike for the pulsed source-sink experiment, the dilution indices of the baseline experiment for $\sigma = 0.1$ do not coincide for later times. This is the effect of the reflective boundary condition, which decreases the effect of diffusion close to the domain boundary.

\section{Conclusion}\label{sec:conclusion}

In this work, we present two contributions. First, we develop an accurate method to adaptively select a grid size for the approximation of an entropy-based measure, the dilution index, preserving its key feature to monotonically increase in time. Our approach allows the dilution index to effectively quantify the enhanced mixing between a solute and a solvent in chaotic flow fields. In the second part, we apply our method to analyze two chaotic injection-extraction systems, revealing that not all chaotic setups lead to very high mixing enhancement.

Evaluating the dilution index to assess the results of particle tracing simulations reveals an undesirable result. The approximated dilution index can yield any value in $(0, E_{\text{max}}]$, depending on the selected grid size, but independent of the underlying particle distribution. Given a sufficient amount of particles, we provide a fixed criterion on the grid size motivated by the theory of representative elementary volumes. Our method is the first approach to sophisticatedly select the grid size of the dilution index to evaluate particle tracking experiments, guaranteeing accurate and unbiased results that show monotonic growth over time. Increasing the number of particles enables the choice of smaller grid sizes and improves the accuracy of the approximation. 

We apply our method to evaluate the temporal increase in mixing of a solute within its solvent due to chaotic advection. To approximate the solution of the advection-diffusion equation, we use lagrangian particle tracking to preserve the fine plume structures due to chaotic advection and minimize numerical diffusion. Our analysis shows that only some configurations of chaotic advection can enhance mixing compared to non-chaotic advection, as KAM islands limit the chaotic region of the chaotic flow. The results demonstrate the fundamental role of diffusion in systems with chaotic advection in enhancing mixing. Diffusion significantly contributes to filling the KAM island and, hence, reaching complete mixing in the systems.
Comparing to a baseline without advection, configurations with advection consistently lead to faster mixing. Our analysis does not consider the effect of the initial solute particle location. In a low diffusivity setup where the solute is initially placed inside a KAM island, the mixing in the chaotic setups can be worse than that of non-chaotic advection. 

Considering chaotic injection-extraction systems for groundwater remediation purposes, our research emphasizes the importance of an analysis of the mixing potential of a system prior to its implementation into the field. Due to high uncertainties in the plume location, a configuration with small KAM islands is preferred to achieve the best remediation results. In order to generate more accurate predictions on the expected mixing enhancement in the field, we envision future research considering heterogeneities in the domain as they would appear due to the geological characteristics of the soil. As it is shown that these soil heterogeneities lead to increased mixing, we expect interactions with the chaotic flow field, which will modify the KAM island structure. To account for the non-Fickian transport in heterogeneous porous media, it would be necessary to adapt our particle tracking method according to \cite{srinivasan_2010}.
The topic of randomizing the flow field is additionally important, as it can further reduce KAM islands. If the flow parameters of the chaotic setups are varied randomly in time, the KAM island structure may disappear, increasing the mixing potential.

\section*{CRediT authorship contribution statement}

\textbf{Carla Feistner:} Conceptualization, Formal analysis, Investigation, Methodology, Software, Visualization, Writing -- original draft, 
\textbf{Mónica Basilio Hazas:} Writing -- review \& editing, 
\textbf{Barbara Wohlmuth:} Supervision, Writing -- review \& editing, 
\textbf{Gabriele Chiogna:} Conceptualization, Funding acquisition, Project administration, Supervision, Writing -- review \& editing

\section*{Declaration of competing interest}

The authors declare that they have no known competing financial interests or personal relationships that could have appeared to influence the work reported in this paper.

\section*{Acknowledgments}

This research is a result of the ChaosAD project (Chaotic ADvection and Mixing Enhancement in Porous Media: The Quest for Experimental Evidence), which is supported by the Deutsche Forschungsgemeinschaft (DFG) under the grant numbers CH 981/8-1 and SI 2853/2-1. We gratefully acknowledge the financial support for B. Wohlmuth partially provided by the DFG under the grant number WO 671/11-1.

\bibliography{references_bibtex}

@article{aref_1984,
  title = {Stirring by Chaotic Advection},
  author = {Aref, Hassan},
  year = {1984},
  month = jun,
  journal = {Journal of Fluid Mechanics},
  volume = {143},
  pages = {1--21},
  issn = {0022-1120, 1469-7645},
  doi = {10.1017/S0022112084001233},
  urldate = {2023-09-18},
  langid = {english}
}

@article{aref_1990,
  title = {Chaotic Advection of Fluid Particles},
  author = {Aref, Hassan},
  year = {1990},
  month = nov,
  journal = {Philosophical Transactions of the Royal Society of London. Series A: Physical and Engineering Sciences},
  volume = {333},
  number = {1631},
  pages = {273--288},
  issn = {0962-8428, 2054-0299},
  doi = {10.1098/rsta.1990.0161},
  urldate = {2023-09-18},
  langid = {english}
}

@article{aref_2002,
  title = {The Development of Chaotic Advection},
  author = {Aref, Hassan},
  year = {2002},
  month = apr,
  journal = {Physics of Fluids},
  volume = {14},
  number = {4},
  pages = {1315--1325},
  doi = {10.1063/1.1458932}
}

@article{aref_2017,
  title = {Frontiers of Chaotic Advection},
  author = {Aref, Hassan and Blake, John R. and Budi{\v s}i{\'c}, Marko and Cardoso, Silvana S. S. and Cartwright, Julyan H. E. and Clercx, Herman J. H. and El Omari, Kamal and Feudel, Ulrike and Golestanian, Ramin and Gouillart, Emmanuelle and {van Heijst}, G. F. and Krasnopolskaya, Tatyana S. and Le Guer, Yves and MacKay, Robert S. and Meleshko, Vyacheslav V. and Metcalfe, Guy and Mezi{\'c}, Igor and {de Moura}, Alessandro P. S. and Piro, Oreste and Speetjens, Michel F. M. and Sturman, Rob and Thiffeault, Jean-Luc and Tuval, Idan},
  year = {2017},
  month = jun,
  journal = {Reviews of Modern Physics},
  volume = {89},
  number = {2},
  pages = {025007},
  issn = {0034-6861, 1539-0756},
  doi = {10.1103/RevModPhys.89.025007},
  urldate = {2023-03-14},
  langid = {english}
}

@article{bagtzoglou_2007,
  title = {Chaotic {{Advection}} and {{Enhanced Groundwater Remediation}}},
  author = {Bagtzoglou, Amvrossios C. and Oates, Peter M.},
  year = {2007},
  month = jan,
  journal = {Journal of Materials in Civil Engineering},
  volume = {19},
  number = {1},
  pages = {75--83},
  issn = {0899-1561, 1943-5533},
  doi = {10.1061/(ASCE)0899-1561(2007)19:1(75)},
  urldate = {2023-10-31},
  langid = {english}
}

@article{basiliohazas_2022,
  title = {Linking Mixing and Flow Topology in Porous Media: {{An}} Experimental Proof},
  shorttitle = {Linking Mixing and Flow Topology in Porous Media},
  author = {Basilio Hazas, M{\'o}nica and Ziliotto, Francesca and Rolle, Massimo and Chiogna, Gabriele},
  year = {2022},
  month = mar,
  journal = {Physical Review E},
  volume = {105},
  number = {3},
  pages = {035105},
  issn = {2470-0045, 2470-0053},
  doi = {10.1103/PhysRevE.105.035105},
  urldate = {2024-01-25},
  langid = {english}
}

@book{bear_1988,
  title = {Dynamics of Fluids in Porous Media},
  author = {Bear, Jacob},
  year = {1988},
  series = {Dover Books on Physics and Chemistry},
  publisher = {Dover},
  address = {New York},
  isbn = {978-0-486-65675-5},
  lccn = {TA357 .B38 1988}
}

@article{borgne_2014,
  title = {Impact of Fluid Deformation on Mixing-induced Chemical Reactions in Heterogeneous Flows},
  author = {Borgne, Tanguy Le and Ginn, Timothy R. and Dentz, Marco},
  year = {2014},
  month = nov,
  journal = {Geophysical Research Letters},
  volume = {41},
  number = {22},
  pages = {7898--7906},
  issn = {0094-8276, 1944-8007},
  doi = {10.1002/2014GL062038},
  urldate = {2024-01-30},
  langid = {english}
}

@book{bremaud_2017,
  title = {Discrete {{Probability Models}} and {{Methods}}},
  author = {Br{\'e}maud, Pierre},
  year = {2017},
  series = {Probability {{Theory}} and {{Stochastic Modelling}}},
  volume = {78},
  publisher = {Springer International Publishing},
  address = {Cham},
  doi = {10.1007/978-3-319-43476-6},
  urldate = {2023-05-29},
  isbn = {978-3-319-43475-9 978-3-319-43476-6},
  langid = {english}
}

@book{chate_1999,
  title = {Mixing: {{Chaos}} and {{Turbulence}}},
  shorttitle = {Mixing},
  editor = {Chat{\'e}, H. and Villermaux, E. and Chomaz, J.-M.},
  year = {1999},
  series = {{{NATO ASI Series}}},
  volume = {373},
  publisher = {Springer US},
  address = {Boston, MA},
  doi = {10.1007/978-1-4615-4697-9},
  urldate = {2023-03-14},
  isbn = {978-1-4613-7127-4 978-1-4615-4697-9},
  langid = {english}
}

@article{cho_2019,
  title = {Field {{Trials}} of {{Chaotic Advection}} to {{Enhance Reagent Delivery}}},
  author = {Cho, Michelle S. and Solano, Felipe and Thomson, Neil R. and Trefry, Michael G. and Lester, Daniel R. and Metcalfe, Guy},
  year = {2019},
  month = jun,
  journal = {Groundwater Monitoring \& Remediation},
  volume = {39},
  number = {3},
  pages = {23--39},
  issn = {1069-3629, 1745-6592},
  doi = {10.1111/gwmr.12339},
  urldate = {2024-01-30},
  langid = {english}
}

@book{durst_2022,
  title = {Fluid {{Mechanics}}: {{An Introduction}} to the {{Theory}} of {{Fluid Flows}}},
  shorttitle = {Fluid {{Mechanics}}},
  author = {Durst, Franz},
  year = {2022},
  series = {Graduate {{Texts}} in {{Physics}}},
  publisher = {Springer Berlin Heidelberg},
  address = {Berlin, Heidelberg},
  doi = {10.1007/978-3-662-63915-3},
  urldate = {2023-08-23},
  isbn = {978-3-662-63913-9 978-3-662-63915-3},
  langid = {english}
}

@article{einstein_1906,
  title = {Zur {{Theorie}} Der {{Brownschen Bewegung}}},
  author = {Einstein, A.},
  year = {1906},
  month = jan,
  journal = {Annalen der Physik},
  volume = {324},
  number = {2},
  pages = {371--381},
  issn = {0003-3804, 1521-3889},
  doi = {10.1002/andp.19063240208},
  urldate = {2024-01-16},
  langid = {english}
}

@article{ganesan_1997,
  title = {Chaotic Heat Transfer Enhancement in Rotating Eccentric Annular-Flow Systems},
  author = {Ganesan, Venkat and Bryden, Michelle D. and Brenner, Howard},
  year = {1997},
  month = may,
  journal = {Physics of Fluids},
  volume = {9},
  number = {5},
  pages = {1296--1306},
  issn = {1070-6631, 1089-7666},
  doi = {10.1063/1.869245},
  urldate = {2024-01-31},
  langid = {english}
}

@article{ghoniem_1985,
  title = {Grid-Free Simulation of Diffusion Using Random Walk Methods},
  author = {Ghoniem, Ahmed F. and Sherman, Frederick S.},
  year = {1985},
  month = oct,
  journal = {Journal of Computational Physics},
  volume = {61},
  number = {1},
  pages = {1--37},
  issn = {00219991},
  doi = {10.1016/0021-9991(85)90058-0},
  urldate = {2023-10-30},
  langid = {english}
}

@article{heyman_2020,
  title = {Stretching and Folding Sustain Microscale Chemical Gradients in Porous Media},
  author = {Heyman, Joris and Lester, Daniel R. and Turuban, R{\'e}gis and M{\'e}heust, Yves and Le Borgne, Tanguy},
  year = {2020},
  month = jun,
  journal = {Proceedings of the National Academy of Sciences},
  volume = {117},
  number = {24},
  pages = {13359--13365},
  issn = {0027-8424, 1091-6490},
  doi = {10.1073/pnas.2002858117},
  urldate = {2024-01-26},
  langid = {english}
}

@article{jones_1988,
  title = {Chaotic Advection in Pulsed Source--Sink Systems},
  author = {Jones, Scott W. and Aref, Hassan},
  year = {1988},
  journal = {Physics of Fluids},
  volume = {31},
  number = {3},
  pages = {469},
  issn = {00319171},
  doi = {10.1063/1.866828},
  urldate = {2023-04-03},
  langid = {english}
}

@article{kadoch_2012,
  title = {A Volume Penalization Method for Incompressible Flows and Scalar Advection--Diffusion with Moving Obstacles},
  author = {Kadoch, Benjamin and Kolomenskiy, Dmitry and Angot, Philippe and Schneider, Kai},
  year = {2012},
  month = jun,
  journal = {Journal of Computational Physics},
  volume = {231},
  number = {12},
  pages = {4365--4383},
  issn = {00219991},
  doi = {10.1016/j.jcp.2012.01.036},
  urldate = {2024-10-22},
  copyright = {https://www.elsevier.com/tdm/userlicense/1.0/},
  langid = {english}
}

@article{kapoor_1998,
  title = {Concentration Fluctuations and Dilution in Aquifers},
  author = {Kapoor, Vivek and Kitanidis, Peter K.},
  year = {1998},
  month = may,
  journal = {Water Resources Research},
  volume = {34},
  number = {5},
  pages = {1181--1193},
  issn = {0043-1397, 1944-7973},
  doi = {10.1029/97WR03608},
  urldate = {2024-01-29},
  langid = {english}
}

@article{kitanidis_1994,
  title = {The {{Concept}} of the {{Dilution Index}}},
  author = {Kitanidis, Peter K},
  year = {1994},
  month = jul,
  journal = {Water Resources Research},
  volume = {30},
  number = {7},
  pages = {2011--2026},
  doi = {10.1029/94WR00762},
  langid = {english}
}

@incollection{krasnopolskaya_2009,
  title = {Quality {{Measures}} and {{Transport Properties}}},
  booktitle = {Analysis and {{Control}} of {{Mixing}} with an {{Application}} to {{Micro}} and {{Macro Flow Processes}}},
  author = {Krasnopolskaya, Tatyana and Meleshko, Vyatcheslav},
  editor = {Maier, Giulio and Salen{\c c}on, Jean and Schneider, Wilhelm and Schrefler, Bernhard and Serafini, Paolo and Cortelezzi, Luca and Mezi{\'c}, Igor},
  year = {2009},
  volume = {510},
  pages = {291--306},
  publisher = {Springer Vienna},
  address = {Vienna},
  doi = {10.1007/978-3-211-99346-0_7},
  urldate = {2024-01-08},
  isbn = {978-3-211-99345-3 978-3-211-99346-0}
}

@article{kree_2017,
  title = {Scalar Mixtures in Porous Media},
  author = {Kree, Mihkel and Villermaux, Emmanuel},
  year = {2017},
  month = oct,
  journal = {Physical Review Fluids},
  volume = {2},
  number = {10},
  pages = {104502},
  issn = {2469-990X},
  doi = {10.1103/PhysRevFluids.2.104502},
  urldate = {2024-01-26},
  langid = {english}
}

@article{leborgne_2008,
  title = {Spatial {{Markov}} Processes for Modeling {{Lagrangian}} Particle Dynamics in Heterogeneous Porous Media},
  author = {Le Borgne, Tanguy and Dentz, Marco and Carrera, Jesus},
  year = {2008},
  month = aug,
  journal = {Physical Review E},
  volume = {78},
  number = {2},
  pages = {026308},
  issn = {1539-3755, 1550-2376},
  doi = {10.1103/PhysRevE.78.026308},
  urldate = {2024-01-25},
  langid = {english}
}

@article{leborgne_2008a,
  title = {Lagrangian {{Statistical Model}} for {{Transport}} in {{Highly Heterogeneous Velocity Fields}}},
  author = {Le Borgne, Tanguy and Dentz, Marco and Carrera, Jesus},
  year = {2008},
  month = aug,
  journal = {Physical Review Letters},
  volume = {101},
  number = {9},
  pages = {090601},
  issn = {0031-9007, 1079-7114},
  doi = {10.1103/PhysRevLett.101.090601},
  urldate = {2024-01-25},
  langid = {english}
}

@article{lemons_1997,
  title = {Paul {{Langevin}}'s 1908 Paper ``{{On}} the {{Theory}} of {{Brownian Motion}}'' [``{{Sur}} La Th{\'e}orie Du Mouvement Brownien,'' {{C}}. {{R}}. {{Acad}}. {{Sci}}. ({{Paris}}) {\textbf{146}} , 530--533 (1908)]},
  author = {Lemons, Don S. and Gythiel, Anthony},
  year = {1997},
  month = nov,
  journal = {American Journal of Physics},
  volume = {65},
  number = {11},
  pages = {1079--1081},
  issn = {0002-9505, 1943-2909},
  doi = {10.1119/1.18725},
  urldate = {2024-01-16},
  langid = {english}
}

@article{lester_2009,
  title = {Lagrangian Topology of a Periodically Reoriented Potential Flow: {{Symmetry}}, Optimization, and Mixing},
  shorttitle = {Lagrangian Topology of a Periodically Reoriented Potential Flow},
  author = {Lester, D. R. and Metcalfe, G. and Trefry, M. G. and Ord, A. and Hobbs, B. and Rudman, M.},
  year = {2009},
  month = sep,
  journal = {Physical Review E},
  volume = {80},
  number = {3},
  pages = {036208},
  issn = {1539-3755, 1550-2376},
  doi = {10.1103/PhysRevE.80.036208},
  urldate = {2023-07-19},
  langid = {english}
}

@article{lester_2013,
  title = {Is {{Chaotic Advection Inherent}} to {{Porous Media Flow}}?},
  author = {Lester, D. R. and Metcalfe, G. and Trefry, M. G.},
  year = {2013},
  month = oct,
  journal = {Physical Review Letters},
  volume = {111},
  number = {17},
  pages = {174101},
  issn = {0031-9007, 1079-7114},
  doi = {10.1103/PhysRevLett.111.174101},
  urldate = {2024-01-26},
  langid = {english}
}

@article{liu_2000,
  title = {Passive Mixing in a Three-Dimensional Serpentine Microchannel},
  author = {Liu, R.H. and Stremler, M.A. and Sharp, K.V. and Olsen, M.G. and Santiago, J.G. and Adrian, R.J. and Aref, H. and Beebe, D.J.},
  year = {2000},
  month = jun,
  journal = {Journal of Microelectromechanical Systems},
  volume = {9},
  number = {2},
  pages = {190--197},
  issn = {1057-7157, 1941-0158},
  doi = {10.1109/84.846699},
  urldate = {2023-09-23}
}

@article{luo_2006,
  title = {Predictions of Quasi-Periodic and Chaotic Motions in Nonlinear {{Hamiltonian}} Systems},
  author = {Luo, Albert C.J.},
  year = {2006},
  month = may,
  journal = {Chaos, Solitons \& Fractals},
  volume = {28},
  number = {3},
  pages = {627--649},
  issn = {09600779},
  doi = {10.1016/j.chaos.2005.08.012},
  urldate = {2024-10-22},
  copyright = {https://www.elsevier.com/tdm/userlicense/1.0/},
  langid = {english}
}

@article{metcalfe_2009,
  title = {Mixing and Heat Transfer of Highly Viscous Food Products with a Continuous Chaotic Duct Flow},
  author = {Metcalfe, Guy and Lester, Daniel},
  year = {2009},
  month = nov,
  journal = {Journal of Food Engineering},
  volume = {95},
  number = {1},
  pages = {21--29},
  issn = {02608774},
  doi = {10.1016/j.jfoodeng.2009.04.032},
  urldate = {2024-01-05},
  langid = {english}
}

@article{morvillo_2021,
  title = {A Scalable Parallel Algorithm for Reactive Particle Tracking},
  author = {Morvillo, Maria and Rizzo, Calogero B. and De Barros, Felipe P.J.},
  year = {2021},
  month = dec,
  journal = {Journal of Computational Physics},
  volume = {446},
  pages = {110664},
  issn = {00219991},
  doi = {10.1016/j.jcp.2021.110664},
  urldate = {2024-10-22},
  langid = {english}
}

@article{neupauer_2014,
  title = {Chaotic Advection and Reaction during Engineered Injection and Extraction in Heterogeneous Porous Media},
  author = {Neupauer, Roseanna M. and Meiss, James D. and Mays, David C.},
  year = {2014},
  month = feb,
  journal = {Water Resources Research},
  volume = {50},
  number = {2},
  pages = {1433--1447},
  issn = {00431397},
  doi = {10.1002/2013WR014057},
  urldate = {2023-12-04},
  langid = {english}
}

@book{ottino_1989,
  title = {The Kinematics of Mixing: Stretching, Chaos, and Transport},
  shorttitle = {The Kinematics of Mixing},
  author = {Ottino, J. M.},
  year = {1989},
  series = {Cambridge Texts in Applied Mathematics},
  number = {3},
  publisher = {Cambridge University Press},
  address = {Cambridge ; New York},
  isbn = {978-0-521-36335-8 978-0-521-36878-0},
  lccn = {TP156.M5 O87 1989}
}

@article{paster_2014,
  title = {Connecting the Dots: {{Semi-analytical}} and Random Walk Numerical Solutions of the Diffusion--Reaction Equation with Stochastic Initial Conditions},
  shorttitle = {Connecting the Dots},
  author = {Paster, Amir and Bolster, Diogo and Benson, David A.},
  year = {2014},
  month = apr,
  journal = {Journal of Computational Physics},
  volume = {263},
  pages = {91--112},
  issn = {00219991},
  doi = {10.1016/j.jcp.2014.01.020},
  urldate = {2024-10-22},
  langid = {english}
}

@article{paster_2015,
  title = {Incomplete Mixing and Reactions in Laminar Shear Flow},
  author = {Paster, A. and Aquino, T. and Bolster, D.},
  year = {2015},
  month = jul,
  journal = {Physical Review E},
  volume = {92},
  number = {1},
  pages = {012922},
  issn = {1539-3755, 1550-2376},
  doi = {10.1103/PhysRevE.92.012922},
  urldate = {2024-01-25},
  langid = {english}
}

@article{peterka_2016,
  title = {Self-{{Adaptive Density Estimation}} of {{Particle Data}}},
  author = {Peterka, Tom and Croubois, Hadrien and Li, Nan and Rangel, Esteban and Cappello, Franck},
  year = {2016},
  month = jan,
  journal = {SIAM Journal on Scientific Computing},
  volume = {38},
  number = {5},
  pages = {S646-S666},
  issn = {1064-8275, 1095-7197},
  doi = {10.1137/15M1016308},
  urldate = {2023-10-30},
  langid = {english}
}

@article{rahbaralam_2015,
  title = {Do We Really Need a Large Number of Particles to Simulate Bimolecular Reactive Transport with Random Walk Methods? {{A}} Kernel Density Estimation Approach},
  shorttitle = {Do We Really Need a Large Number of Particles to Simulate Bimolecular Reactive Transport with Random Walk Methods?},
  author = {Rahbaralam, Maryam and {Fern{\`a}ndez-Garcia}, Daniel and {Sanchez-Vila}, Xavier},
  year = {2015},
  month = dec,
  journal = {Journal of Computational Physics},
  volume = {303},
  pages = {95--104},
  issn = {00219991},
  doi = {10.1016/j.jcp.2015.09.030},
  urldate = {2024-10-22},
  langid = {english}
}

@article{rolle_2019,
  title = {Mixing and {{Reactive Fronts}} in the {{Subsurface}}},
  author = {Rolle, Massimo and Le Borgne, Tanguy},
  year = {2019},
  month = sep,
  journal = {Reviews in Mineralogy and Geochemistry},
  volume = {85},
  number = {1},
  pages = {111--142},
  issn = {1529-6466},
  doi = {10.2138/rmg.2018.85.5},
  urldate = {2024-01-29},
  langid = {english}
}

@article{salamon_2006,
  title = {A Review and Numerical Assessment of the Random Walk Particle Tracking Method},
  author = {Salamon, Peter and {Fern{\`a}ndez-Garcia}, Daniel and {G{\'o}mez-Hern{\'a}ndez}, J. Jaime},
  year = {2006},
  month = oct,
  journal = {Journal of Contaminant Hydrology},
  volume = {87},
  number = {3-4},
  pages = {277--305},
  issn = {01697722},
  doi = {10.1016/j.jconhyd.2006.05.005},
  urldate = {2024-10-23},
  copyright = {https://www.elsevier.com/tdm/userlicense/1.0/},
  langid = {english}
}

@article{schelin_2009,
  title = {Chaotic Advection in Blood Flow},
  author = {Schelin, A. B. and K{\'a}rolyi, {\relax Gy}. and {de Moura}, A. P. S. and Booth, N. A. and Grebogi, C.},
  year = {2009},
  month = jul,
  journal = {Physical Review E},
  volume = {80},
  number = {1},
  pages = {016213},
  issn = {1539-3755, 1550-2376},
  doi = {10.1103/PhysRevE.80.016213},
  urldate = {2024-01-26},
  langid = {english}
}

@article{schmidt_2017,
  title = {A {{Kernel-based Lagrangian}} Method for Imperfectly-Mixed Chemical Reactions},
  author = {Schmidt, Michael J. and Pankavich, Stephen and Benson, David A.},
  year = {2017},
  month = may,
  journal = {Journal of Computational Physics},
  volume = {336},
  pages = {288--307},
  issn = {00219991},
  doi = {10.1016/j.jcp.2017.02.012},
  urldate = {2024-10-22},
  langid = {english}
}

@article{song_2003,
  title = {A {{Microfluidic System}} for {{Controlling Reaction Networks}} in {{Time}}},
  author = {Song, Helen and Tice, Joshua D. and Ismagilov, Rustem F.},
  year = {2003},
  month = feb,
  journal = {Angewandte Chemie International Edition},
  volume = {42},
  number = {7},
  pages = {768--772},
  issn = {1433-7851, 1521-3773},
  doi = {10.1002/anie.200390203},
  urldate = {2024-01-31},
  langid = {english}
}

@article{souzy_2020,
  title = {Velocity Distributions, Dispersion and Stretching in Three-Dimensional Porous Media},
  author = {Souzy, M. and Lhuissier, H. and M{\'e}heust, Y. and Le~Borgne, T. and Metzger, B.},
  year = {2020},
  month = may,
  journal = {Journal of Fluid Mechanics},
  volume = {891},
  pages = {A16},
  issn = {0022-1120, 1469-7645},
  doi = {10.1017/jfm.2020.113},
  urldate = {2024-01-26},
  langid = {english}
}

@article{sposito_1998,
  title = {Tracer Advection by Steady Groundwater Flow in a Stratified Aquifer},
  author = {Sposito, Garrison and Weeks, Scott W.},
  year = {1998},
  month = may,
  journal = {Water Resources Research},
  volume = {34},
  number = {5},
  pages = {1051--1059},
  issn = {00431397},
  doi = {10.1029/98WR00009},
  urldate = {2023-04-28},
  langid = {english}
}

@article{sposito_2006,
  title = {Chaotic Solute Advection by Unsteady Groundwater Flow},
  shorttitle = {Chaotic Solute Advection by Unsteady Groundwater Flow},
  author = {Sposito, Garrison},
  year = {2006},
  month = jun,
  journal = {Water Resources Research},
  volume = {42},
  number = {6},
  issn = {00431397},
  doi = {10.1029/2005WR004518},
  urldate = {2023-03-14},
  langid = {english}
}

@article{srinivasan_2010,
  title = {Random Walk Particle Tracking Simulations of Non-{{Fickian}} Transport in Heterogeneous Media},
  author = {Srinivasan, G. and Tartakovsky, D.M. and Dentz, M. and Viswanathan, H. and Berkowitz, B. and Robinson, B.A.},
  year = {2010},
  month = jun,
  journal = {Journal of Computational Physics},
  volume = {229},
  number = {11},
  pages = {4304--4314},
  issn = {00219991},
  doi = {10.1016/j.jcp.2010.02.014},
  urldate = {2024-10-22},
  copyright = {https://www.elsevier.com/tdm/userlicense/1.0/},
  langid = {english}
}

@article{stone_2001,
  title = {Microfluidics: {{Basic}} Issues, Applications, and Challenges},
  shorttitle = {Microfluidics},
  author = {Stone, H. A. and Kim, S.},
  year = {2001},
  month = jun,
  journal = {AIChE Journal},
  volume = {47},
  number = {6},
  pages = {1250--1254},
  issn = {00011541, 15475905},
  doi = {10.1002/aic.690470602},
  urldate = {2023-09-23},
  langid = {english}
}

@article{stremler_2004,
  title = {Designing for Chaos: Applications of Chaotic Advection at the Microscale},
  shorttitle = {Designing for Chaos},
  author = {Stremler, Mark A. and Haselton, F. R. and Aref, Hassan},
  editor = {Ottino, J. M. and Wiggins, S. R.},
  year = {2004},
  month = may,
  journal = {Philosophical Transactions of the Royal Society of London. Series A: Mathematical, Physical and Engineering Sciences},
  volume = {362},
  number = {1818},
  pages = {1019--1036},
  issn = {1364-503X, 1471-2962},
  doi = {10.1098/rsta.2003.1360},
  urldate = {2023-03-14},
  langid = {english}
}

@article{tartakovsky_2008,
  title = {Stochastic {{Langevin Model}} for {{Flow}} and {{Transport}} in {{Porous Media}}},
  author = {Tartakovsky, Alexandre M. and Tartakovsky, Daniel M. and Meakin, Paul},
  year = {2008},
  month = jul,
  journal = {Physical Review Letters},
  volume = {101},
  number = {4},
  pages = {044502},
  issn = {0031-9007, 1079-7114},
  doi = {10.1103/PhysRevLett.101.044502},
  urldate = {2024-01-25},
  langid = {english}
}

@article{tartakovsky_2010,
  title = {Langevin Model for Reactive Transport in Porous Media},
  author = {Tartakovsky, Alexandre M.},
  year = {2010},
  month = aug,
  journal = {Physical Review E},
  volume = {82},
  number = {2},
  pages = {026302},
  issn = {1539-3755, 1550-2376},
  doi = {10.1103/PhysRevE.82.026302},
  urldate = {2024-01-25},
  langid = {english}
}

@article{teh_2008,
  title = {Droplet Microfluidics},
  author = {Teh, Shia-Yen and Lin, Robert and Hung, Lung-Hsin and Lee, Abraham P.},
  year = {2008},
  journal = {Lab on a Chip},
  volume = {8},
  number = {2},
  pages = {198},
  issn = {1473-0197, 1473-0189},
  doi = {10.1039/b715524g},
  urldate = {2024-01-05},
  langid = {english}
}

@book{tome_2015,
  title = {Stochastic {{Dynamics}} and {{Irreversibility}}},
  author = {Tom{\'e}, T{\^a}nia and De Oliveira, M{\'a}rio J.},
  year = {2015},
  series = {Graduate {{Texts}} in {{Physics}}},
  publisher = {Springer International Publishing},
  address = {Cham},
  doi = {10.1007/978-3-319-11770-6},
  urldate = {2023-05-10},
  isbn = {978-3-319-11769-0 978-3-319-11770-6},
  langid = {english}
}

@article{turuban_2018,
  title = {Space-{{Group Symmetries Generate Chaotic Fluid Advection}} in {{Crystalline Granular Media}}},
  author = {Turuban, R. and Lester, D. R. and Le Borgne, T. and M{\'e}heust, Y.},
  year = {2018},
  month = jan,
  journal = {Physical Review Letters},
  volume = {120},
  number = {2},
  pages = {024501},
  issn = {0031-9007, 1079-7114},
  doi = {10.1103/PhysRevLett.120.024501},
  urldate = {2024-01-26},
  langid = {english}
}

@article{ward_2015,
  title = {Mixing in Microfluidic Devices and Enhancement Methods},
  author = {Ward, Kevin and Fan, Z Hugh},
  year = {2015},
  month = sep,
  journal = {Journal of Micromechanics and Microengineering},
  volume = {25},
  number = {9},
  pages = {094001},
  issn = {0960-1317, 1361-6439},
  doi = {10.1088/0960-1317/25/9/094001},
  urldate = {2024-01-31}
}

@article{warhaft_2000,
  title = {Passive {{Scalars}} in {{Turbulent Flows}}},
  author = {Warhaft, Z.},
  year = {2000},
  month = jan,
  journal = {Annual Review of Fluid Mechanics},
  volume = {32},
  number = {1},
  pages = {203--240},
  issn = {0066-4189, 1545-4479},
  doi = {10.1146/annurev.fluid.32.1.203},
  urldate = {2024-01-26},
  langid = {english}
}

@article{weeks_1998,
  title = {Mixing and Stretching Efficiency in Steady and Unsteady Groundwater Flows},
  author = {Weeks, Scott W. and Sposito, Garrison},
  year = {1998},
  month = dec,
  journal = {Water Resources Research},
  volume = {34},
  number = {12},
  pages = {3315--3322},
  issn = {0043-1397, 1944-7973},
  doi = {10.1029/98WR02535},
  urldate = {2024-01-26},
  langid = {english}
}

@article{ye_2015,
  title = {Experimental {{Evidence}} of {{Helical Flow}} in {{Porous Media}}},
  author = {Ye, Yu and Chiogna, Gabriele and Cirpka, Olaf A. and Grathwohl, Peter and Rolle, Massimo},
  year = {2015},
  month = nov,
  journal = {Physical Review Letters},
  volume = {115},
  number = {19},
  pages = {194502},
  issn = {0031-9007, 1079-7114},
  doi = {10.1103/PhysRevLett.115.194502},
  urldate = {2024-01-30},
  langid = {english}
}

@article{ye_2016,
  title = {Experimental Investigation of Transverse Mixing in Porous Media under Helical Flow Conditions},
  author = {Ye, Yu and Chiogna, Gabriele and Cirpka, Olaf A. and Grathwohl, Peter and Rolle, Massimo},
  year = {2016},
  month = jul,
  journal = {Physical Review E},
  volume = {94},
  number = {1},
  pages = {013113},
  issn = {2470-0045, 2470-0053},
  doi = {10.1103/PhysRevE.94.013113},
  urldate = {2024-01-25},
  langid = {english}
}

@article{younes_2005,
  title = {Solving the Advection-Diffusion Equation with the {{Eulerian}}--{{Lagrangian}} Localized Adjoint Method on Unstructured Meshes and Non Uniform Time Stepping},
  author = {Younes, A. and Ackerer, P.},
  year = {2005},
  month = sep,
  journal = {Journal of Computational Physics},
  volume = {208},
  number = {1},
  pages = {384--402},
  issn = {00219991},
  doi = {10.1016/j.jcp.2005.02.019},
  urldate = {2024-10-22},
  copyright = {https://www.elsevier.com/tdm/userlicense/1.0/},
  langid = {english}
}

\appendix

\section{Analytical solution of the advective process in the pulsed source-sink system}\label{app:PSS_analytical_sol}

The analytical solution of the flow defined in \eqref{eq:PSS_hamiltonian_system} is derived in \cite{sposito_2006,jones_1988}. It is shown that due to the operation of a point sink/source a particle does move on a straight line towards/away from the sink/source. Thereby its squared distance to the sink/source over time $t$ is reduced/increased by $\lambda^2(t) := Qt/pi$. Hence, while operating the sink for time $t$, all particles within the circle of radius $\lambda(t)$ are extracted from the domain. Vise versa, by operating the source for time $t$, the newly injected particles fill the circle with radius $\lambda(t)$ around the source as indicated in \cref{tikz:exp_setup}.
For a source at location $\textmatrix{0\\0}$ we can define the resulting flow by
\begin{align}\label{eq:PSS_basic_flow}
  \Psi^t_+\bigmatrix{x\\y} 
  = \bigmatrix{x\\y} \sqrt{1 + \frac{\lambda^2}{x^2+y^2}}.
\end{align}
In the case of a sink at the same location $+ \lambda^2$ is replaced by $- \lambda^2$. 
Assuming now arbitrary locations for sink and source, we can convert this configuration to a setup with sink and source at $\pm\textmatrix{a\\0}$ by using a coordinate transformation. Multiplying the length scale by $1/a$ and using $\Lambda = \lambda/a$, we can further assume sink and source to be located at $\pm\textmatrix{1\\0}$, without loss of generality \cite{jones_1988}.
Given a source at location $\textmatrix{-1\\0}$ the flow in \eqref{eq:PSS_basic_flow} is modified to
\begin{align*}
  \Psi^t_+\bigmatrix{x\\y}  = 
  \bigmatrix{x + 1\\y} \,
  \sqrt{1 + \frac{\Lambda^2(t)}{(x + 1)^2 + y^2}} - \bigmatrix{1\\0}.
\end{align*}
In our setup, sink and source operate alternatively for time $\tau$, using $\Lambda := \Lambda(\tau)$, we define the map $M_+:\R^2 \rightarrow \R^2$ for operating the source and $M_-:\R^2 / B_\Lambda\textmatrix{1\\0} \rightarrow \R^2$ for operating the sink and map $\textmatrix{x(t)\\y(t)} \mapsto \textmatrix{x(t+\tau)\\y(t+\tau)}$.

For the reinjection procedure, we find that each particle $\textmatrix{x\\y}$ within the circle of radius $\Lambda$ around $\textmatrix{1\\0}$, i.e., $\textmatrix{x\\y} \in B_\Lambda\textmatrix{1\\0}$, is extracted from the domain after time $p\tau$ ($p \in [0, 1]$). Following the \textit{first-out-first-in} protocol the same particle is injected by the source at time $p\tau$ after the source started operating. The total time the particle moves with the flow remains $\tau$ and hence the distance to the source after its full stroke must be $\sqrt{\Lambda^2 - \left(x^2 + y^2\right)}$. We additionally assume, the injection angle to be $\pi - \theta$ with $\theta$ being the extraction angle. This leads to the map $M_{\text{inj}}:B_\Lambda\textmatrix{1\\0} \rightarrow \R^2$
\begin{align*}
  M_{\text{inj}}\bigmatrix{x\\y} = \bigmatrix{1 - x\\y}\sqrt{\frac{\Lambda^2}{(x-1)^2+y^2} - 1} - \bigmatrix{1\\0}.
\end{align*}
Combining these expressions, we construct the map $M:\R^2 \rightarrow \R^2$ representing the consecutive operation of source and sink where the sink is operated first:
\begin{align*}
  M\bigmatrix{x\\y} = \begin{cases}
    \left(M_+ \circ M_-\right)\textmatrix{x\\y} & \text{for } \textmatrix{x\\y} \in \R^2/B_\Lambda\textmatrix{1\\0} \\
    M_{\text{inj}}\textmatrix{x\\y} & \text{otherwise}.
  \end{cases}
\end{align*}

\section{Analytical solution to the advective process in the reference system of the RPM flow}\label{app:RPM_analytical}

\begin{figure}[!tbp!]
  \centering
  \includegraphics[width=8.6cm]{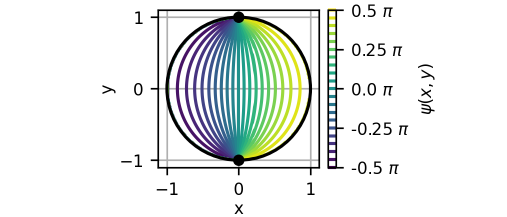}
  
  \caption{Contour lines of the stream function $\psi$ of the RPM flow.}
  \label{fig:RPM_contours}
\end{figure} 

\begin{figure}[!tbp!]
  \centering
  \includegraphics[width=8.6cm]{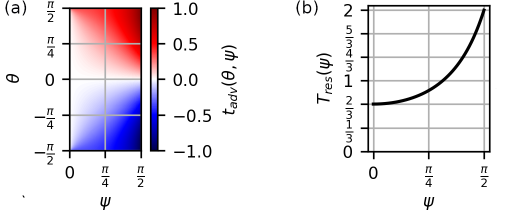}
  
  \caption{(a) Advection time $t_{\text{adv}}$ and (b) residence time $T_{\text{res}}$ over their respective parameter spaces.}
  \label{fig:RPM_t_adv_T_res}
\end{figure} 

The analytical solution of Hamiltonian system of the reference system is derived in \cite{lester_2009}. Thereby source and sink are fixed at locations $\pm\textmatrix{0\\1}$ while the rotation of the RPM flow is implemented by rotating the particles. As the flow is symmetrical to the $y$-axis, we only respect the positive side of the $x$-axis. The authors of \cite{lester_2009} utilize the fact that the streamlines of the flow are given by the contour lines of the Hamiltonian $\psi$, plotted in \cref{fig:RPM_contours}. The position of each particle can be represented uniquely by its angle to the origin $\theta \in (-\pi/2, \pi/2)$ and its streamline $\psi \in (0, \pi/2]$. A representation in this coordinate system is convenient as for each particle $\psi$ is constant in time. They establish an expression for the advection time of a particle along a streamline until it reaches $\theta = 0$. It is given by
\begin{align}\label{eq:RPM_t_adv_app}
  \begin{split}
    t_\text{adv}(\theta, \psi) = \csc^2(\psi)
    \times\Biggl\{ &\cot(\psi) \arctan \left[ \frac{\sin(\theta) \cot(\psi)}{\sqrt{1 + \cos^2(\theta) \cot^2(\psi)}} \right] \\
    &+ \sin(\theta) \sqrt{1 + \cos^2(\theta) \cot^2(\psi)}
    - |\cot(\psi)| (\theta + \cos(\theta) \sin(\theta)) \Biggl\}.
    \end{split}
\end{align}
The expression uses the cosecant defined as $\csc(x)=\frac{1}{\sin(x)}$ and the cotangent given by $\cot(x) = \frac{1}{\tan(x)}.$
For particles downstream of the centerline $\theta = 0$ this yields a negative advection time. The residence time of a particle injected by the source onto streamline $\psi$ before it is extracted by the sink can be directly computed from \eqref{eq:RPM_t_adv}, as the sum of the advection time from the source to the centerline plus the advection time from the centerline to the sink:
\begin{align}\label{eq:RPM_t_res_app}
  T_{\text{res}}(\psi) = \left(t_{\text{adv}}\left(\frac{\pi}{2}, \psi\right) - t_{\text{adv}}\left(-\frac{\pi}{2}, \psi\right)\right).
\end{align}
Both quantities are plotted in \cref{fig:RPM_t_adv_T_res}.
Combining \eqref{eq:RPM_t_adv_app} and \eqref{eq:RPM_t_res_app}, the travel time of a particle until it reaches the sink is given by $t_{\text{adv}}(\theta, \psi) + T_{\text{res}}(\psi)/2$. If a particle does not get extracted within time $t$ it holds that 
\begin{align*}
  t_{\text{adv}}(\theta(t), \psi_0) + \frac{T_{\text{res}}(\psi_0)}{2} = t_{\text{adv}}(\theta_0, \psi_0) + \frac{T_{\text{res}}(\psi_0)}{2} - t.
\end{align*}
Respecting that a particle extracted by the sink is injected instantaneously by the source onto the same streamline, the above equation is extended to
\begin{align}\label{eq:RPM_formular_advection_app}
  \begin{split}
    t_{\text{adv}}(\theta(t), \psi_0) + \frac{T_{\text{res}}(\psi_0)}{2} = 
    \left(t_{\text{adv}}(\theta_0, \psi_0) + \frac{T_{\text{res}}(\psi_0)}{2} - t\right) \mod T_{\text{res}}(\psi_0).
  \end{split}
\end{align}
Thereby, the modulo operator $\mod$ takes care of the extraction reinjection process. If the travel time of a particle $(\theta_0, \psi_0)$ to the sink is smaller than $t$, i.e., $t_{\text{adv}}(\theta_0, \psi_0) + T_{\text{res}}(\psi_0)/2 < t$, it gets extracted by the sink. The reinjection by the source is performed instantaneously onto the same streamline. The new travel time to the sink is given by $\left(t_{\text{adv}}(\theta_0, \psi_0) + T_{\text{res}}(\psi_0)/2 - t\right) \mod T_{\text{res}}(\psi_0)$. We remark that the corresponding equation in \cite{lester_2009} uses a plus sign in front of the $t$ on the right-hand side. As we can reproduce the experimental results in \cite{lester_2009} using the corrected version of the equation given in \eqref{eq:RPM_formular_advection_app}, we can show that the sign change is only a typo in the formula. The numerical results of \cite{lester_2009} are still valid. 
Solving \eqref{eq:RPM_formular_advection_app} yields the flow $\Psi_1^{t}: \{\textmatrix{x\\y}\in B_1\textmatrix{0\\0}: x > 0\} \rightarrow \{\textmatrix{x\\y}\in B_1\textmatrix{0\\0}: x > 0\}$ which can be used for all particles on the positive side of the $x$-axis. For particles on the negative side, we use the reflection symmetry of $\psi$ with respect to the $y$-axis, yielding the flow $\Psi_2^{t}: \{\textmatrix{x\\y}\in B_1\textmatrix{0\\0}: x < 0\} \rightarrow \{\textmatrix{x\\y}\in B_1\textmatrix{0\\0}: x < 0\}$, $\Psi_2^{t}\textmatrix{x\\y} = \textmatrix{-1&0\\0&1}\Psi_1^{t}\textmatrix{-x\\y}$. 

For particles on the $y$-axis (i.e., $x(t) = 0$), we can use the description given in \cref{sec:CA_setups}. As the usage of $(\theta, \psi)$ coordinates is insufficient, we switch to Cartesian coordinates, yielding
\begin{align*}
  t_{\text{adv}}(y) = \frac{y}{2} - \frac{y^3}{6}, \qquad T_{\text{res}} = \frac{2}{3},
\end{align*}
which can be plugged into \eqref{eq:RPM_formular_advection_app} where $y$ takes the place of $\theta$.
We denote the resulting flow on the $y$-axis by $\Psi_0^t: \{\textmatrix{x\\y}\in \R^2: x=0\} \rightarrow \{\textmatrix{x\\y}\in \R^2: x=0\}$.
Combining the above flows, we are able to define the flow $\Psi^t:B_1\textmatrix{0\\0} \rightarrow B_1\textmatrix{0\\0}$ for the whole domain by
\begin{align*}
  \Psi^t\bigmatrix{x\\y} = \begin{cases}
    \Psi_0^t\textmatrix{x\\y} & \text{if } ~ x = 0 \\
    \Psi_1^t\textmatrix{x\\y} & \text{if } ~ x > 0 \\
    \Psi_2^t\textmatrix{x\\y} & \text{if } ~ x < 0.
  \end{cases}
\end{align*}

\pagebreak

\clearpage
\begin{center}
  {\LARGE Supplementary material}
\end{center}

\renewcommand{\thesection}{S\arabic{section}}
\renewcommand{\thefigure}{S\arabic{figure}}
\renewcommand{\thetable}{S\arabic{table}}
\setcounter{table}{0}
\setcounter{figure}{0}
\setcounter{section}{0}

\section{Introduction}

The supporting material shows our numerical results in more details. For the motivation of our adaptive approach of selecting the dilution index we show numerical results also for different amounts of particles, highlighting how the approximation error can be decreased by simultaneously increasing the amount of particles and decreasing the grid size in section \cref{sec:app_gaussian}. For the numerical experiment concerning mixing enhancement of the RPM flow we show the results for the dilution index, the growth rate and the convergence rate, for all considered parameter combinations, in section \cref{sec:app_RPM}. 

\section{A computationally reliable approach for the approximation of the dilution index}\label{sec:app_gaussian}

The usage of an increased number of particles in the particle tracking experiments enables the method to select smaller grid sizes for the dilution index approximation. Similar to Fig. 6 in the paper we created \cref{fig:gaussian}. Plot (a) shows the dilution index for different values of the diffusion coefficient $\sigma$ and different amounts of particles $m$. We show the upper limit $E \leq m\Delta V$ for the approximated dilution index for all considered choices of $m$. For a lager $m$ we find the plateau in \cref{fig:gaussian}(a) to be elongated compared to smaller $m$. We also show the relative approximation error in \cref{fig:gaussian}(b) and $\Delta_\delta[E](h)$ in \cref{fig:gaussian}(c). We find that, as expected, the approximation error of the dilution index $E$ reaches lower values the more particles are used in the particle tracking experiment. The observations made in the paper for $m = 10\,000$ are also valid for different choices of $m$.

\section{Numerical results for the RPM flow}\label{sec:app_RPM}

In the paper we concentrate on the evolution of the reactor ratio $M$ over time $t$ for flow parametrizations $\left(\Theta, \tau\right) \in \left\{\left(0, 0.2\right), \left(\pi/6, 0.2\right), \left(\pi/6, 0.5\right), \left(\pi/2, 0.2\right), \left(\pi/2, 0.5\right)\right\}$. The graph including all considered flow parametrizations is given in \cref{fig:RPM_convergence} and \cref{fig:RPM_dil_over_t_loglog}. We observe very different results depending on the different choices of the flow parametrization, due to the very individual KAM island structure.

\begin{figure}[tbp!]
  \centering
  \includegraphics[width=8.6cm]{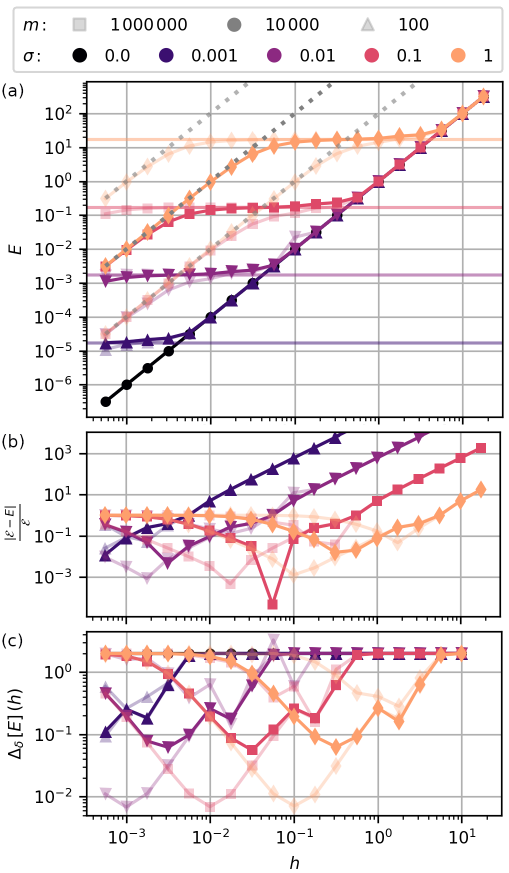}
  \caption{(a) Shows the value of the dilution index $E$ over the grid size $h$ for different diffusion coefficients $\sigma$ and numbers of particles $m$. The dotted line gives the upper bound of the dilution index $m\Delta V$. 
  The continuous lightened lines help to visualize the plateaus forming at the value of the analytical dilution index.
  (b) Shows the relative error $\frac{|\mathcal{E} - E|}{\mathcal{E}}$ of the approximated dilution index $E$ depending on the grid size $h$ to the analytical value $\mathcal{E}$. We observe a convergence $E \rightarrow \mathcal{E}$, until we pass the optimal grid size $h^*$. For smaller $h$, the dilution index dissipates from the analytical value.
  (c) Shows the approximated derivative of the dilution index $\Delta_\delta[E](h)$, using forward finite differences in the log-log scale, for $\sigma \neq 0$. The minima of the derivatives correspond approximately to the minima in the approximation error of plot (b).}
  \label{fig:gaussian}
\end{figure}

\begin{figure*}[tbp!]
  \centering
  \includegraphics[width=15.2cm]{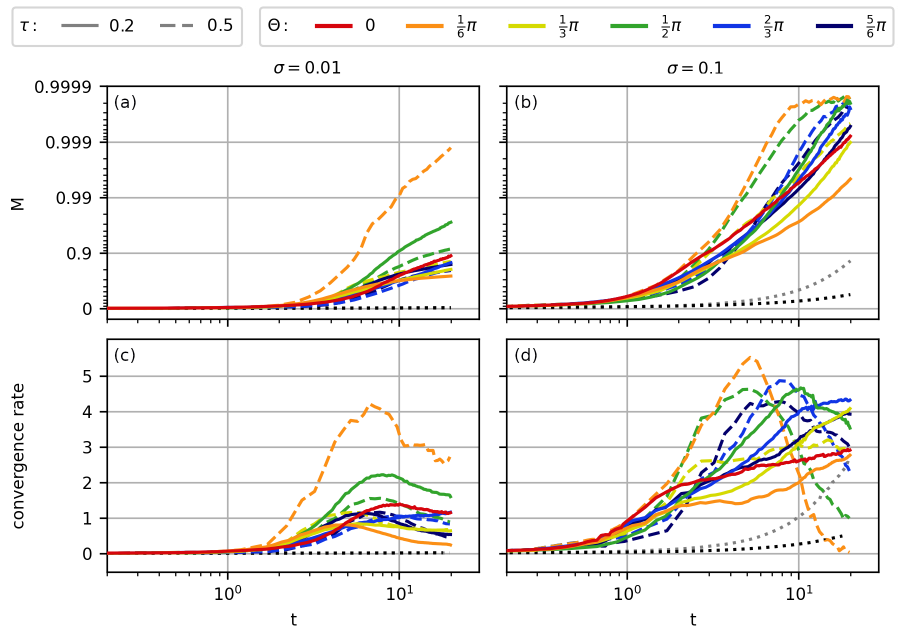}

  \caption{Convergence of the reactor ration $M$ to $M_{\text{max}}=1$. (a)-(b) show the evolution of $M$ over $t$ for a logarithmic typed $y$-axis highlighting the convergence $M \rightarrow 1$, (c)-(d) show the convergence rate for $M$ approaching $1$ for different flow parametrization. The convergence rate was computed using the moving average over the first order finite difference by using a window of $\pm 0.2$ in the exponent. The dotted lines correspond to the case without advection where the initial particle position is either at one point (gray) or in a ring (black).}
  \label{fig:RPM_convergence}
\end{figure*}

\begin{figure*}[tbp!]
  \centering
  \includegraphics[width=15.2cm]{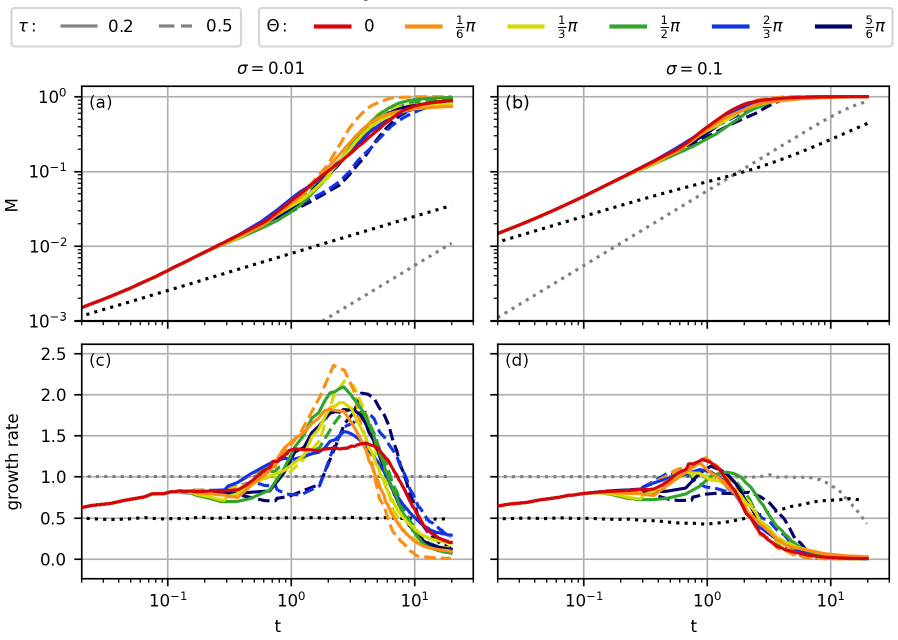}

  \caption{(a)-(b) reactor ratio $M$ and (c)-(d) reactor ratio growth rate over time $t$ for different flow parametrization. The growth rate was approximated using the moving average over the first order finite differences with a window of $\pm 0.2$ in the exponent. The dotted lines correspond to the case without advection where the initial particle position is either at one point (gray) or in a ring (black).}
  \label{fig:RPM_dil_over_t_loglog}
\end{figure*}

\end{document}